\documentclass[12pt]{article}

\usepackage{fullpage}
\usepackage{amsmath, amsthm, amssymb}
\usepackage{graphicx}
\usepackage{enumerate}
\usepackage{natbib}
\usepackage{subfig}
\usepackage{multirow}
\usepackage{float}
\usepackage{tikz}
\usepackage{longtable}
\usepackage{indentfirst}
\usepackage{setspace}
\RequirePackage[colorlinks,citecolor=blue,urlcolor=blue]{hyperref}

\newtheorem{lemma}{Lemma}

\newtheorem{theorem}{Theorem}

\DeclareMathOperator{\diag}{diag}

\DeclareMathOperator*{\argmin}{argmin}
\DeclareMathOperator{\RCV}{RCV}
\DeclareMathOperator{\MSE}{MSE}

\definecolor{c1}{rgb}{0,  0, 1}
\definecolor{c2}{rgb}{1,  0, 0}

\DeclareRobustCommand\full  {\tikz[baseline=-0.6ex]\draw[c1, thick] (0,0)--(0.57,0);} 
\DeclareRobustCommand\denselydashed{\tikz[baseline=-0.6ex]\draw[c2, thick, dash pattern={on 7pt off 1.5pt}] (0,0)--(0.57,0);} 

\begin{document}

\title{Robust penalized estimators for functional linear regression}
\author{Ioannis Kalogridis and Stefan Van Aelst}
\date{%
    Department of Mathematics, KU Leuven, Belgium\\[2ex]%
    \today
}

\maketitle

\begin{abstract}
Functional data analysis is a fast evolving branch of statistics. Estimation procedures for the popular functional linear model either suffer from lack of robustness or are computationally burdensome. To address these shortcomings, a flexible family of penalized lower-rank estimators based on a bounded loss function is proposed. The proposed class of estimators is shown to be consistent and can attain high rates of convergence with respect to prediction error under weak regularity conditions. These results can be generalized to higher dimensions under similar assumptions. The finite-sample performance of the proposed family of estimators is investigated by a Monte-Carlo study which shows that these estimators reach high efficiency while offering protection against outliers. The proposed estimators compare favorably to existing approaches robust as well as non-robust alternatives. The good performance of the method is also illustrated on a complex real dataset.
\end{abstract}



\section{Introduction}
\label{sec:intro}

In recent years, technological innovations and improved storage capabilities have led practitioners to observe and record increasingly complex high-dimensional data. Among others, data that are characterized by an underlying functional structure have attracted considerable research interest, following works such as \citet{Ramsay:1982}, \citet{Ramsay:1991} and \citet{Ramsay:2005}. Particular interest has been devoted to the functional linear model, relating a scalar response $Y$ to a random function $X$, which is viewed as an element of $(\Omega, \mathcal{A}, \mathcal{P})$ with sample paths in $\mathcal{L}^2(\mathcal{I})$,  through the model
\begin{equation}
\label{eq:1}
Y = \alpha_0 + \int_{\mathcal{I}} X(t) f_0(t) dt + \sigma_0\epsilon.
\end{equation}
Here, $\alpha_0 \in \mathbb{R}$ is the intercept, $f_0$ is a square integrable coefficient (weight) function defined on a compact interval $\mathcal{I}$ of a Euclidean space, $\sigma_0$ is an unknown scale parameter and $\epsilon$ is a random error, that is assumed to be independent of $X$. Typically, $\epsilon$ is also assumed to possess finite second moments, but this assumption is not needed for the theoretical results in this paper.

The vast domain of applications of the model, ranging from  meteorology~\citep{Ramsay:2005} and chemometrics \citep{Ferraty:2006} to diffusion tensor imaging tractography \citep{GS:2014}, has spurred the development of numerous novel estimation methods.  Since estimating the coefficient function $\beta$ is an infinite dimensional problem, regularization through dimension reduction or penalization is crucial for the success of these methods. Regressing on the scores of the leading functional principal components \citep{cardot1999functional} is the oldest and perhaps to this day the most popular method of estimation. However, although consistent \citep{Hall:2007}, functional principal component regression may fail to yield smooth estimates of the coefficient function, even in moderately large samples. This fact has motivated proposals that explicitly impose smoothness of the estimated coefficient function. \citet{Cardot:2003} proposed estimation through a penalized spline expansion while functional extensions of smoothing splines have been proposed and studied by \citet{Crambes:2009} and \citet{Yuan:2010}. A hybrid approach between principal component and penalized spline regression  has been developed by \citet{reiss2007functional} and \citet{GS:2011}, who combine these methods in order to attain greater flexibility. 

Variable selection ideas have also been adapted to the functional regression setting. \citet{James:2009} proposed  imposing sparsity on higher order derivatives of a high dimensional basis expansion of $\beta$ in order to produce more interpretable estimates. Expressing the coefficient function in the wavelet domain, \citet{Zhao:2012} proposed an $\ell_1$ regularization scheme in order to select the most relevant resolutions and ensure stable and accurate estimates of a wide variety of coefficient functions. For more details on existing estimation methods as well as informative comparisons, one may consult the comprehensive review papers  of \citet{Morris:2015} and \citet{reiss2017methods}.

Since all of the above methods rely on generalized least-squares type estimators, a drawback in their use is that the presence of outliers can have a serious effect on the resulting estimates. 
To address this lack of robustness, more robust estimation procedures have been introduced.
\citet{Maronna:2013} proposed a robust version of the smoothing spline estimator of \citet{Crambes:2009} but did not study theoretical properties of their method. \citet{Shin:2016} have extended the work of \citet{Yuan:2010} by considering more outlier-resistant loss functions and showed that under regularity conditions their M-type smoothing spline estimator attains the same rates of convergence as its least-squares counterpart. Similarly, \citet{Qingguo:2017} generalized the work of \citet{Hall:2007} to functional principal component regression with a general convex loss function. More recently, \citet{Boente:2020} proposed a family of sieves estimators based on bounded loss functions and B-spline expansions and investigated rates of convergence with respect to the prediction error. 

In general, sieves estimators based on either functional principal components or B-splines and smoothing spline estimators can be considered to be situated on the two ends of a spectrum. Unpenalized sieves estimators are easy to implement, yet frequently result in either undersmoothed or oversmoothed estimates of the regression function.  This undesirable feature results from the discrete nature of their smoothing parameter, which in this case is the dimension of the basis.  On the other hand, smoothing spline estimators, while capable of yielding estimates with the right amount of smoothness, can be unwieldy due to their high dimension. In particular, the requirement to have as many basis functions as the sample size leads to computationally challenging estimators that are prone to instabilities due to the often complex nature of functional data. In the nonparametric regression framework, the case for lower-rank representations on the grounds of simplicity has already been made by \citet{Wahba:1990}. For functional regression, an even stronger case can be made due to the lack of banded matrices that enable fast computational algorithms for smoothing splines in this setting. 

As a compromise between these two types of estimators, this paper introduces and studies a family of lower-rank penalized estimators based on the principle of MM-estimation, as described by \citet{Yohai:1987}. The proposed class of estimators exhibit a high degree of robustness against both vertical outliers and leverage points, while also maintaining high efficiency under Gaussian errors. In our opinion, this class of estimators fills an important void in the literature by providing a family of flexible and resistant estimators that is also computationally feasible. Our framework does not only include the popular B-spline basis combined with a quadratic roughness penalty, but also many other basis systems combined with a wealth of possible penalties. Examples include the Fourier basis with the harmonic acceleration penalty introduced by \citet{Ramsay:2005} and the wavelet basis with bounded variation or Besov penalties \citep[Chapter 10]{van de Geer:2000}. 

The remainder of the paper is organized as follows. Section \ref{sec:2} introduces the proposed  family of penalized estimators and discusses some popular choices of basis systems and penalties in more detail. Section \ref{sec:3}  is devoted to the study of the asymptotic properties of these estimators. We show that under mild regularity conditions the estimators achieve a high rate of convergence with respect to the commonly considered prediction error.  Our regularity conditions do not require the existence of any moments of the error term, allowing in effect for very heavy-tailed error distributions. Our analysis also uncovers a useful error decomposition pointing to the roles of the variance as well as the twin biases stemming from modelling and regularization. Sections \ref{sec:4}  and \ref{sec:5} illustrate the competitive finite-sample performance of the proposed estimator in a Monte Carlo study and in real data. Section \ref{sec:6} contains a final discussion while all proofs are collected in the appendix.

\section{Robust penalized estimators for functional linear regression}
\label{sec:2}

\subsection{Penalized MM-estimators with general bases and penalties}

Let us consider independent and identically distributed tuples $(X_1, Y_1), \ldots, (X_n, Y_n)$ which satisfy model \eqref{eq:1}. For simplicity we shall identify $\mathcal{I}$ with $[0,1]$, without loss of generality. A popular estimation approach for the functional linear model \citep[Chapter 15]{Ramsay:2005} expands the functional slope $\beta$ in terms of a dense set of $\mathcal{L}^2([0,1])$ functions $\{f_i\}_i$, then truncates this expansion and finally estimates the coefficients using a roughness penalty. Let $||\cdot||$ and $\langle \cdot, \cdot \rangle$ denote the usual $\mathcal{L}^2([0,1])$ norm and inner product, respectively. Moreover, let $\Theta_K$ denote the $K$-dimensional linear subspace of $\mathcal{L}^2([0,1])$ spanned by $f_1, \ldots, f_K$, then this strategy amounts to solving
\begin{equation}
\label{eq:2}
(\widehat{\alpha}_{LS}, \widehat{f}_{LS} ) = \argmin_{\alpha \in \mathbb{R}, f \in \Theta_K } \left[ \frac{1}{n} \sum_{i=1}^n \left| Y_i - \alpha - \langle X_i, f \rangle \right|^2 + \lambda ||f^{(q)}||^2. \right].
\end{equation}
Hence, the roughness penalty is placed on the integrated squared $q$th derivative of $f$ and it is weighted by a penalty parameter $\lambda \geq 0$, which is usually chosen in a data-driven way.  The penalty parameter places a premium on the roughness of the estimated function as measured by its integrated squared $q$th derivative. Large values of $\lambda$ force the estimated coefficient function to behave essentially like a polynomial of degree at most $q-1$ while small values of $\lambda$ produce more wiggly estimates. It is important to note that for such estimators regularization is accomplished by both restricting the basis functions ($f \in \Theta_K$) and penalizing roughness. This strategy leads to more complex estimators than unpenalized sieve estimators  but considerably less complex estimators than the smoothing spline estimators of \citet{Crambes:2009, Yuan:2010} and \citet{Shin:2016}.

It is well-known that the least-squares criterion employed in \eqref{eq:2} yields estimators that are susceptible to outlying observations. To protect against such anomalies we propose to replace the square loss function by a bounded loss function $\rho$ and estimate the unknown quantities according to
\begin{align}
\label{eq:3}
(\widehat{\alpha}_n, \widehat{f}_n ) = \argmin_{\alpha \in \mathbb{R}, f \in \Theta_{K} } \left[  \frac{1}{n} \sum_{i=1}^n \rho \left( \frac{Y_i - a - \langle X_i, f \rangle}{\widehat{\sigma}_n} \right)  + \lambda \mathcal{J}(f) \right],
\end{align}
where $\widehat{\sigma}_n$ is a robust estimator of the scale of the error and $\mathcal{J}(f): \Theta_K \to \mathbb{R}_{+}$ is a general penalty functional on  $\Theta_K$, usually a seminorm. For $\lambda = 0$ the penalty term vanishes and we obtain an unpenalized sieve estimator, such as the B-spline estimator proposed by \citet{Boente:2020}. On the other hand, for $\lambda \to \infty$ the penalty will dominate the objective function and forces the estimator to lie in the null-space of $\mathcal{J}(\cdot)$. The present set-up is very general and allows for a wide variety of approximating subspaces and penalties. To illustrate this flexibility, we now discuss three important examples of basis systems and penalties that are permitted within our framework.

\noindent
\textbf{Example 1} (B-splines with derivative or difference penalties): Fix an integer $p \geq 1$, select $t_{1}< \ldots< t_K$ distinct locations within $(0,1)$ and define the spline subspace
\begin{align*}
\Theta_{K+p} = \left\{f: f(x) = \sum_{j=1}^{K+p} f_j B_{j,p}(x) \right\},
\end{align*}
where $B_{j,p}, j = 1, \ldots, K+p$, are the B-splines of order $p$ supported by $t_1 \ldots, t_K,$ with $2p$ arbitrary boundary knots. For $p=1$, $\Theta_{K+p}$ consists of all step functions with jumps at the knots $t_{i}$ while for $p \geq 2$, $\Theta_{K+p}$ is a subspace of $\mathcal{C}^{p-2}([0,1])$ with the property that each $f \in \Theta_{K+p}$ is a polynomial of order $p$ on each subinterval $[t_i, t_{i+1}]$. The common choice $\mathcal{J}(f) = ||f^{(q)}||^2$ for some integer $q<p$ was introduced by \citet{O:1986}. Another popular choice is the P-spline penalty \citep{Eilers:1996}, given by
\begin{align*}
\mathcal{J}(f) = \sum_{j=q+1}^{K+p} |\Delta^q f_j|^2, 
\end{align*}
where $\Delta^q$ refers to the $q$th-order backward difference operator. This difference penalty largely retains the mathematical properties of the derivative penalty, but results in much simpler expressions.  In the frequently used setting of equidistant knots, the derivative and difference penalties on spline subspaces are scaled versions of one another, see, e.g., Proposition 1 of \citet{Kalogridis:2021a}.

\noindent
\textbf{Example 2} (Fourier expansion with derivative or harmonic acceleration penalties). Consider the trigonometric sieve given by
\begin{align*}
\Theta_{K} = \left\{f: f(x) = \alpha_0 +  \sum_{j=1}^{K}\{\alpha_j \cos(2 \pi j x) + \beta_j \sin(2 \pi j x)\}\right\}.
\end{align*}
This sieve consists of infinitely differentiable functions with increasing amplitude. Unlike the B-spline basis, the Fourier basis is not local, but it is orthonormal and its derivatives are orthogonal, resulting in simple expressions. For instance, taking $\mathcal{J}(f) = ||f^{(q)}||^2$, as in \citet{Li:2007}, leads to $\mathcal{J}(f) = \mathbf{f}^{\top} \mathbf{D} \mathbf{f}$ with  $\mathbf{f} = (\boldsymbol{\alpha}, \boldsymbol{\beta})$ and
\begin{align*}
\mathbf{D} = \diag\{ (2 \pi)^{2q}, (2\pi)^{2q}, (4\pi)^{2q}, (4\pi)^{2q}, \ldots, (2\pi K)^{2q} \}.
\end{align*}
Another possibility is the harmonic acceleration penalty proposed in \citet[Chapter 15]{Ramsay:2005} which is given by $\mathcal{J}(f) = \int_{0}^1 \left| (4 \pi^2)f^{\prime}(x) + f^{(3)}(x) \right|^2 dx$.
Interestingly, this penalty shrinks the solution towards a function of the form $f(x) = \alpha_0 + \alpha_1 \sin(2\pi x) + \beta_1 \cos(2\pi x)$.

\noindent
\textbf{Example 3} (Wavelets with total variation or $\ell_1$ penalties). Choose a scaling function $\phi$ and a mother wavelet $\psi$ that are orthonormal in $\mathcal{L}^2([0,1])$. Now, for $(j,k) \in \mathbb{N} \times \mathbb{N}$ put $\phi_{j,k}(x) = 2^{j/2} \phi(2^j x-k)$ and $\psi_{j,k}(x) = 2^{j/2} \psi(2^j x-k)$. Fix $j_0 \in \mathbb{N}$ and set $J = \log_2 K -1$ for $K \in \mathbb{N}$. The wavelet subspace with primary decomposition level $j_0$ is given by
\begin{align*}
\Theta_K = \left\{f: f(x) = \sum_{k = 0}^{2^{j_0}-1} f^{\prime}_{j_0,k} \phi_{j_0,k}(x) + \sum_{j=j_0}^{J} \sum_{k=0}^{2^{j}-1} f_{j,k} \psi_{j,k}(x)  \right\}.
\end{align*}
This wavelet subspace involves $2^{J+1} = K$ coefficients. Possible penalties are the total variation penalty with $q=1$, i.e., $\mathcal{J}(f) = \int_{0}^1 |f^{\prime}(x)| dx$ or the $\ell_1$ penalty on all the coefficients given by
\begin{align*}
\mathcal{J}(f) = \sum_{k = 0}^{2^{j_0}-1} |f^{\prime}_{j_0,k}| + \sum_{j=j_0}^{J} \sum_{k=0}^{2^{j}-1} |f_{j,k}|,
\end{align*}
as used by \citet{Zhao:2012} in the context of least-squares estimation. 

A widely used family of bounded smooth $\rho$-functions that is useful for our purposes is the family of Tukey bisquare loss functions, defined as
\begin{align*}
\rho_c(x) = \begin{cases} 1-\left\{1-(x/c)^2 \right\}^3 & |x| \leq c \\
1 & |x| >c, \end{cases},
\end{align*}
where $c>0$ is a tuning parameter that determines the trade-off between robustness and efficiency 
\citep{Maronna:2019}. In the central part, that is, for $|x| \leq c$, $\rho$ the loss function is strictly increasing and it smoothly transitions to a constant function as $|x| \to c$. Thus, the loss incurred by large residuals is constant leading to regression estimators that are impervious to large outliers. 

The scale estimate $\widehat{\sigma}_n$ is an important part of the estimator, as it essentially acts as an additional tuning parameter for the loss function. A robust scale estimate may be obtained from an M-scale of the residuals of an S-estimator \citep{Rousseeuw:1984}. In particular, for $\alpha \in \mathbb{R}$ and $f \in \Theta_K$ let $\widehat{\sigma} _n= \widehat{\sigma}_n(\mathbf{r})$ be an M-scale estimate based on a vector of residuals
\begin{align*}
\mathbf{r}(\alpha, f) = (r_1(\alpha, f), \ldots, r_n(\alpha, f)),
\end{align*}
with $r_i(a,f) = Y_i - \alpha - \langle X_i, f \rangle$, $i = 1\ldots, n$. Then an S-estimator $(\widehat{\alpha}^{S}, \widehat{f}^{S})$ is defined as 
\begin{align}\
\label{eq:4}
(\widehat{\alpha}^{S}, \widehat{f}^{S}) = \argmin_{\alpha \in \mathbb{R}, f \in \Theta_K} \widehat{\sigma}(\mathbf{r}(\alpha, f)).
\end{align}
We set $\widehat{\sigma}_n$ equal to the S-scale estimate, which is given by the minimum of the objective function in \eqref{eq:4}, i.e., $\widehat{\sigma}_n = \widehat{\sigma}(\mathbf{r}(\alpha^S, f^S))$. Note that S-estimators are well-defined in our setting as $K<n$, i.e., there are fewer parameters than observations. This will also be a requirement for our asymptotic results, see Section 3 below. Hence, computationally efficient algorithms, such as the fast-S algorithm proposed by \citet{SB:2006}, can be applied to obtain the solution of \eqref{eq:4}. 

\subsection{Computational aspects}

The penalized MM-estimator in \eqref{eq:3} depends on the choice of the approximating subspace, its dimension and the penalty parameter. In this section, we outline a number of possible strategies for their selection, but first we briefly discuss the computation of penalized MM-estimates. To this end, we need to differentiate between quadratic and non-quadratic penalties. For quadratic penalties, that is, for penalties which can be written as $\mathcal{J}(f) = \mathbf{f}^{\top} \mathbf{D} \mathbf{f}$ for some positive semi-definite $\mathbf{D}$, a fast computational procedure may be developed along the lines of the penalized variant of iteratively reweighted least-squares given in \citet{Maronna:2011}. To better guarantee that the algorithm returns a global minimum, we recommend initiating the iterations from the robust unpenalized S-estimate given by \eqref{eq:4}. For non-quadratic penalties, such as the $\ell_1$-penalty for instance, we recommend the use of the iterative LARS algorithm proposed by \citet{Smucler:2017}, again starting from the unpenalized S-estimate. 

Let us now consider the choices that need to be made for the penalized MM-estimator. The dimension $K$ of the subspace seems to be the least critical for the success of the estimator. Indeed, extensive experience with lower-rank penalized estimators \citep{Ruppert:2003, Wood:2017} has shown that the dimension does not make much difference for the resulting solution as long as the approximating subspace is rich enough, but $K$ is still smaller than the sample size $n$. In our experience, a choice such as $K = [\min\{40, n/4\}]$, which ensures at least $4$ observations per basis function and puts a cap at $40$ basis functions, is appropriate for many situations. The number of basis functions can be increased beyond 40 in highly complex situations, but these tend to be rather rare in practice.

For the choice of the basis system some guidelines already exist in the literature. For instance, \citet{Ramsay:2005} recommend using the Fourier system for periodic data and the B-spline system otherwise. As we shall see in Section 3, both systems require smoothness of the coefficient function $f_0$, in order to attain high rates of convergence. For cases in which the regression function is suspected to be less smooth, possibly with local characteristics, such as spikes, one may opt for the wavelet system instead. However, in this case special attention must be devoted not only to the tuning parameter $\lambda$, but also to the level of the decomposition $j_0$. As this is a discrete parameter the additional computational burden is not excessive. 

To determine the penalty parameter $\lambda$ in a data-driven way, we propose to select the value of $\lambda$ that minimizes the robust cross-validation (RCV) criterion
\begin{equation*}
\RCV(\lambda) =  \sum_{i=1}^n \frac{   W(r_{i}) |r_{i}|^2 }{{(1- h_{i} )^2 }}  ,
\end{equation*}
where $r_{i}$ is the $i$th residual, $W(r_{i}) = \rho^{\prime}(r_{i})/r_{i}$ and the $h_{i}$ are measures of the influence of the $i$th observation, which can, for instance, be obtained from the diagonal of the weighted hat-matrix obtained upon convergence of the iterative reweighted least-squares algorithm. The $\RCV$ criterion adopted herein may be viewed as a robustification of the classical leave-one-out criterion \citep[see, e.g.,][]{Wahba:1990} in which all $W(r_i)$ are identically equal to one and hence there is no downweighting of outlying observations.

For the simulation experiments and real-data examples in this paper we have adopted a two-step approach to identify the minimizer of $\RCV(\lambda)$. First, we have
determined the approximate location of the minimizer by evaluating $\RCV(\lambda)$ on a grid and then employed a numerical optimizer based on golden section search and parabolic interpolation \citep{Nocedal:2006} in the neighborhood of this approximate optimum. Such a hybrid approach is often advisable due to the possible local minima and near-flat regions of the CV criterion. Implementations and illustrative examples of the penalized MM-estimator may be found in \url{https://github.com/ioanniskalogridis/Robust-functional-linear-regression}.

\section{Asymptotic properties}
\label{sec:3}

\subsection{Consistency}

We now study asymptotic properties of the penalized MM-estimators defined in Section~\ref{sec:2}. For notational convenience we assume that the variables are centred so that $\alpha_0 = 0$ and the object of interest is the coefficient function $f_0$. As is common for spline estimators in the functional linear model, see, e.g., \citet{Cardot:2003, Crambes:2009} and \citet{Boente:2020}, we focus on the distance criterion given by
\begin{align*}
\pi(f, g) = [\mathbb{E} \{| \langle X, f-g \rangle|^2 \} ]^{1/2}, \quad (f, g) \in \mathcal{L}^2([0,1]) \times \mathcal{L}^2([0,1]),
\end{align*}
which may be rewritten as $\pi(f, g) = \{\langle \Gamma (f-g), f-g \rangle\}^{1/2}$ with $\Gamma$ denoting the self-adjoint Hilbert-Schmidt covariance operator of $X$. This criterion is directly linked to the average squared prediction error that arises when using $\langle X_{n+1}, \widehat{f}_n \rangle$ to predict $\langle X_{n+1}, f_0 \rangle$, where $X_{n+1}$ is a new random function possessing the same distribution as $X$.

For our theoretical development we require assumptions on the loss function, $\rho$, the error, $\epsilon$, the functional predictor, $X$, and the coefficient function, $f_0$. For each $K$-dimensional approximating subspace $\Theta_K$ we define the element closest to $f_0$ as $\widetilde{f}_K = \argmin_{f \in \Theta_K} ||f_0-f||$. Since $\Theta_K$ is closed and convex, the Hilbert projection theorem ensures that $\widetilde{f}_K$ is a well-defined and unique element of $\Theta_K$. Note that $\widetilde{f}_K$ is an abstract quantity to which we have no access in practice, but its existence and properties are essential for the results to follow. We require the following assumptions.

\begin{itemize}
\item[(A1)] The loss function $\rho$ satisfies $\rho(0)=0$ and is even, non-decreasing on $[0,\infty)$, bounded and twice continuously differentiable with bounded derivatives $\psi$ and $\psi^{\prime}$. Furthermore, $\sup_{x \in \mathbb{R}}|x\psi(x)| < \infty$. Without loss of generality we assume that $||\rho||_{\infty} = 1$.
\item[(A2)] The scale estimate $\widehat{\sigma}_n$ satisfies $\widehat{\sigma}_n \xrightarrow{P} \sigma_0$, where $\sigma_0$ is defined in \eqref{eq:1}.
\item[(A3)] The error $\epsilon$ is independent of $X$ and possesses a Lebesgue-density $g_0(t)$ that is even, decreasing in $|t|$ and strictly decreasing in $|t|$ in a neighbourhood of zero. Furthermore, $\mathbb{E}\{ \psi^{\prime}(\epsilon)\} >0$.
\item[(A4)] There exists a $C>0$ such that $\Pr(||X||\leq C)= 1$ and for every $f \in \mathcal{L}^2([0,1])$ and $\alpha \in \mathbb{R}$ such that $(f, \alpha) \neq (0,0)$, $\Pr( \langle X, f \rangle) = \alpha)<1$.
\item[(A5)] The coefficient function $f_0$ belongs to a Banach space of functions, $\mathcal{B}([0,1])$, that is embeddable in $\mathcal{C}([0,1])$. Furthermore, the unit ball $\{f \in \mathcal{B}([0,1]): ||f||_{\mathcal{B}} \leq 1 \}$ is compact in the topology of the norm $||\cdot||_{\infty}$.
\item[(A6)] There exists a $c \in (0,1)$ such that $\Pr(\langle X, f\rangle  = 0 )<c$ for any $f \in \mathcal{B}([0,1])$ such that $f \neq 0$.
\item[(A7)] $\Theta_K \subset \mathcal{B}([0,1])$ and the dimension $K$ satisfies $K \asymp n^{\beta}$ for some $\beta \in (0,1)$. Furthermore, $||\widetilde{f}_K-f_0|| \to 0$ as $K \to \infty$ and $\lambda \mathcal{J}(\widetilde{f}_K) \xrightarrow{P} 0$, as $n \to \infty$.
\end{itemize}

Assumptions (A1)--(A3) are standard for MM-estimators, see \citet{Yohai:1987}. In combination with (A6) they imply that the estimators are Fisher-consistent so that at the population level we are indeed estimating the target function $f_0$, see Lemma~\ref{Lem1} in the appendix. Assumption (A1) is satisfied, for example, by the Tukey bisquare loss. As shown by \citet{Boente:2020}, the S-scale estimator  satisfies (A2) under mild conditions. The first part of (A4) imposes the almost sure boundedness of the functional covariate when viewed as an element of $\mathcal{L}^2([0,1])$. This assumption has been used extensively in the asymptotics of the functional linear regression model, see for example \citet{Cardot:2003, Zhao:2012} and \citet{Boente:2020}. The second part of (A4) ensures that $X$ is not concentrated on any subspace of $\mathcal{L}^2([0,1])$, which is the case whenever $X$ possesses a Karhunen-Loève decomposition consisting of infinitely many non-zero terms \citep[Chapter 7]{Hsing:2015}. Equivalently, the null-space of its covariance operator $\Gamma$ should only consist of the zero element.

Assumptions (A5) and (A7) are mild smoothness conditions on the coefficient function. For $\mathcal{B}([0,1])$ to be embeddable in $\mathcal{C}([0,1])$ it suffices to have $\mathcal{B}([0,1]) \subset \mathcal{C}([0,1])$ and a constant $c_0>0$ such that 
\begin{align}
\label{eq:5}
||f||_{\infty} \leq c_0 ||f||_{\mathcal{B}}, \quad f \in \mathcal{B}([0,1]).
\end{align}
Equivalently, the identity operator between these two spaces should be bounded. Furthermore, the unit ball in $\mathcal{B}([0,1])$ should be compact, when merged with $\mathcal{C}([0,1])$. Both parts of assumption (A5) are satisfied by many interesting spaces of functions. Consider, for example, the Sobolev space $\mathcal{W}^{1,p}([0,1])$ defined as
\begin{align*}
\mathcal{W}^{1,p}([0,1]) = \left\{f:[0,1] \to \mathbb{R}: \left\{\int_{0}^1 |f(x)|^p dx\right\}^{1/p} + \left\{\int_{0}^1 | f^{\prime}(x)|^p dx\right\}^{1/p} < \infty \right\},
\end{align*}
with $p>1$. It can be shown that $\mathcal{W}^{1,p}([0,1])$ is complete when endowed with the norm $||f||_{\mathcal{W}^{1,p}} = ||f||_p + ||f^{\prime}||_p$. The mean-value theorem and Hölder's inequality may be employed to show that \eqref{eq:5} holds, while the unit ball $\{f \in \mathcal{W}^{1,p}([0,1]) : ||f||_{W^{1,p}} \leq 1 \}$ is compact in the sup-norm by virtue of the Arzel\`{a}-Ascoli theorem, as this set of functions is equicontinuous. These observations may be naturally extended to higher-order Sobolev spaces, see \citet{Adams:2003}.

Finally, assumption (A7) states that $f_0$ may be arbitrarily well-approximated by an element of $\Theta_K$ in the $\mathcal{L}^2([0,1])$-norm when $K \to \infty$. This approximating sequence $\widetilde{f}_K$ should have finite roughness, as measured by $\mathcal{J}(\cdot)$, so that $\lambda \mathcal{J}(\widetilde{f}_K) \xrightarrow{P} 0$ as $n \to \infty$. In many cases we have $\mathcal{J}(\widetilde{f}_K) = O(1)$, hence if $\lambda \xrightarrow{P} 0$ the assumption is satisfied. It is important to note that in this work we treat $\lambda$ as a random quantity and not merely as a deterministic sequence, as is often the case in literature \citep{Cardot:2003,Yuan:2010, Shin:2016}. In our opinion, this constitutes an important generalization, as in most cases $\lambda$ is selected by a data-driven procedure and thus is random rather than fixed.

Our first result extends Theorem 3.1 of \citet{Boente:2020} for the unpenalized B-spline estimator to the more general setting considered herein. It ensures that the penalized sieve estimators converge uniformly to the target coefficient function $f_0$. By (A4), uniform convergence also implies convergence with respect to prediction error.
\begin{theorem}
\label{Thm:1}
Suppose that assumptions (A1)--(A7) hold. Furthermore, let 
\begin{align*}
M(f, \sigma) = \mathbb{E}\left\{ \rho\left( \frac{Y- \langle X, f \rangle }{\sigma} \right)  \right\},
\end{align*}
and assume that $M(f_0, \sigma_0) =b <1$ and $c <1-b$ in (A6). Then, $||\widehat{f}_n - f_0||_{\mathcal{B}} = O_P(1)$ and  $||\widehat{f}_n-f_0||_{\infty}  = o_P(1)$, as $n \to \infty$. 
\end{theorem}
\noindent
The condition $M(f_0, \sigma_0) =b <1$ required by Theorem~\ref{Thm:1} serves to avoid boundary solutions, in which  $|\epsilon/\sigma_0|$ is so large that $\rho(\epsilon/\sigma_0) = 1$ almost surely (recall that $\rho$ is even and $||\rho||_{\infty}=1$). The second condition $c<1-b$ parallels condition (A3) in \citet{Yohai:1987} and may be viewed as a compatibility condition, see also \citet{Smucler:2017} for a similar use.

\subsection{Rates of convergence} \label{sec:convrate}

The result of Theorem~\ref{Thm:1} covers estimators based on many different basis systems and penalties, which all converge under suitable assumptions. To illustrate potential differences among estimators, we go one step further and investigate their respective rates of convergence in Theorem~\ref{Thm:2} below, which is based on the following development. We begin by defining the finite-sample version  of $M(f,\sigma)$, that is,
\begin{align*}
M_n(f, \sigma) = \frac{1}{n} \sum_{i=1}^n \rho \left( \frac{Y_i - \langle X_i, f \rangle}{\sigma} \right).
\end{align*}
Our objective function is $M_n(f, \widehat{\sigma}_n) + \lambda \mathcal{J}(f)$ and the minimization is over all $f \in \Theta_K$. By the projection theorem, $\widetilde{f}_K \in \Theta_K$ and therefore
\begin{align}\
\label{eq:6}
M_n(\widehat{f}_n, \widehat{\sigma}_n) + \lambda \mathcal{J}(\widehat{f}_n) \leq M_n(\widetilde{f}_K, \widehat{\sigma}_n) + \lambda \mathcal{J}(\widetilde{f}_K).
\end{align}
Adding $M(\widehat{f}_n, \widehat{\sigma}_n)-M(\widetilde{f}_K, \widehat{\sigma}_n)$ on both sides of \eqref{eq:6}, moving $M_n(\widehat{f}_n, \widehat{\sigma}_n)$ to the right-hand side and noting that $\lambda \mathcal{J}(\widehat{f}_n) \geq 0$ yields
\begin{align}
\label{eq:7}
M(\widehat{f}_n, \widehat{\sigma}_n) - M(\widetilde{f}_K, \widehat{\sigma}_n) & \leq U_n( \widetilde{f}_K, \widehat{f}_n,\widehat{\sigma}_n)+ \lambda J(\widetilde{f}_K),
\end{align}
where $U_n(f, g, \sigma) : \Theta_K \times \Theta_K \times \mathbb{R}_{+} \to \mathbb{R}$ denotes the mean-centered process given by
\begin{align*}
U_n(f, g, \sigma) = \{M_n(f, \sigma) - M(f,\sigma)\} +  \{M_n(g, \sigma ) - M(g, \sigma)\}.
\end{align*}
Now, under our assumptions it can be shown that there exist strictly positive constants $\eta$ and $L$ such that
\begin{align}
\label{eq:8}
M(\widehat{f}_n, \widehat{\sigma}_n) - M(\widetilde{f}_K, \widehat{\sigma}_n) \geq \eta | \pi(\widehat{f}_n, \widetilde{f}_K)|^2 - L  ||\widetilde{f}_K - f_0|| \pi(\widehat{f}_n, \widetilde{f}_K),
\end{align}
for all large $n$ with high probability. The regularity of the process $U_n(f,g,\sigma)$ determines the asymptotic variance, cf. Lemma 3.2 in \citet{van de Geer:2002}. In particular, we show that
\begin{align}
\label{eq:9}
U_n(\widetilde{f}_K, \widehat{f}_n,  \widehat{\sigma}_n) = O_{\Pr}(1) \{\gamma_n \pi(\widehat{f}_n, \widetilde{f}_K) \vee \gamma_n^2\}, 
\end{align}
where $\gamma_n = K \log n/n$ and $a \vee b = \max(a,b)$. Rearranging, we obtain
\begin{align}
\label{eq:10}
\eta |\pi(\widehat{f}_n, \widetilde{f}_K)|^2 \leq O_{\Pr}(1) \{\gamma_n \pi(\widehat{f}_n, \widetilde{f}_K) \vee \gamma_n^2\} + O_{\Pr}(1) ||\widetilde{f}_K - f_0|| \pi(\widehat{f}_n, \widetilde{f}_K)  + \lambda \mathcal{J}(\widetilde{f}_{K}).
\end{align}
This inequality involving the square of $\pi(\widehat{f}_n, \widetilde{f}_K)$ in the left-hand side and $\pi(\widehat{f}_n, \widetilde{f}_K)$ in the right-hand side is key to Theorem~\ref{Thm:2} below, see the appendix for a detailed derivation.
\begin{theorem}
\label{Thm:2}
Suppose that assumptions (A1)--(A7) hold, $M(f_0, \sigma_0) < b$ and $c<1-b$ in (A6). Then,
\begin{align*}
|\pi(\widehat{f}_n, f_0)|^2 = O_{\Pr} \left( \frac{K \log n}{n} \right) + O_{\Pr}(||\widetilde{f}_K-f_0||
^2) + O_{\Pr}(\lambda \mathcal{J}(\widetilde{f}_K)),
\end{align*}
as $n \to \infty$.
\end{theorem}

Theorem~\ref{Thm:2} presents the prediction error as a decomposition into three terms, which represent the variance, the modelling bias and the regularization bias respectively.  The variance term depends only on the dimension of the sieve and not on its type. This situation has a well-known parallel in non-parametric regression, \citep[see, e.g.,][Chapter 15]{Eg:2009}. The $\log n$-term appearing in our decomposition is non-standard and results from slightly imprecise local entropy calculations \citep[cf.][Chapter 9]{van de Geer:2000}. An intuitive explanation is that it reflects the difficulty of inference whenever the predictor variable is an infinite-dimensional object.

The second term in the decomposition of Theorem~\ref{Thm:2} is the bias stemming from the approximation of a generic $\mathcal{L}^2([0,1])$-function with a $\Theta_K$-function. To ensure that this approximation error decreases fast as $K = K_n \to \infty$, we need to select a sieve that approximates well the class of functions to which $f_0$ belongs. Lastly, the penalization bias is reflected by the term $\lambda \mathcal{J}(\widetilde{f}_K)$. This term suggests that to obtain high rates of convergence other than an appropriate basis system, one also needs an appropriate measure of roughness $\mathcal{J}(\cdot)$ on $\Theta_K$. For example, given $\lambda$ selecting the wavelet subspace of Example 3 combined with $J(f) = ||f^{(q)}||^2$ would most likely lead to large values of $\mathcal{J}(\widetilde{f}_K)$ thereby diminishing the asymptotic performance of the estimator. Let us now revisit the previous examples and see how the prediction error behaves for some standard choices of $\mathcal{J}(\cdot)$.

\noindent
\textbf{Example 1} (Cont.) Assume that $f_0$ has uniformly bounded derivatives up to order $r \geq 1$ with $r$th derivative satisfying a Lipschitz condition of order $v \in (0,1)$. Note that this space of functions also satisfies (A5) under its usual norm. Then, for $p>r$ and equidistant interior knots we find $||\widetilde{f}_K-f_0||  = O(K^{-r-v)})$,
 \citep[see][p.149]{DB:2001}. At the same time, $||\widetilde{f}^{(q)}||^2 = O(1)$ for all $q<p$ \citep[see, e.g.,][]{Kalogridis:2021a} leading to
\begin{align*}
|\pi(\widehat{f}_n, f_0)|^2 = O_{\Pr} \left( \frac{K \log n}{n} \right) + O_{\Pr}\left(\frac{1}{K^{2r+2v}}\right)+ O_{\Pr}(\lambda).
\end{align*}
For $K \asymp n^{1/\{2(r+v)+1\}}$ and $\lambda =  O_{\Pr}(n^{-\gamma})$ with $\gamma \geq 2(r+v)/ \{2(r+v)+1\}$ we obtain $|\pi(\widehat{f}_n, f_0)|^2 = O_P(n^{-2(r+v)/\{2(r+v)+1\}} \log n)$. This is a much higher rate of convergence than the $n^{-2(r+v)/\{4(r+v)+1\}}$-rate \sloppy obtained by \citet{Cardot:2003} for the penalized least squares estimator, which is a consequence of our use of modern empirical process methodology to derive the result.

\noindent
\textbf{Example 2} (Cont.) Under similar assumptions on $f_0$ as in the previous example we have $||\widetilde{f}_K-f_0||  = O(K^{-r-v)})$, as seen from \citet[Corollary 7.2.4]{Devore:1993}. At the same time [Theorem 7.2.7 of \citet{Devore:1993} implies $||\widetilde{f}^{(q)}||^2 = O(1)$ for $q \leq r$, whence
\begin{align*}
|\pi(\widehat{f}_n, f_0)|^2 = O_{\Pr} \left( \frac{K \log n}{n} \right) + O_{\Pr}\left(\frac{1}{K^{2r+2v}}\right) + O_{\Pr}(\lambda).
\end{align*}
For similar choices of $K$ and $\lambda$ as in the spline setting, we are again led to $|\pi(\widehat{f}_n, f_0)|^2 = O_P(n^{-2(r+v)/\{2(r+v)+1\}} \log n)$. The same conclusion holds for the harmonic acceleration penalty, provided that $r \geq 3$. The fact that many different sieves yield exactly the same rate of convergence for smooth functions is well-known in classical nonparametric regression \citep{Shen:1994}.

\noindent
\textbf{Example 3} (Cont.) For a demonstration of a different flavour consider the Sobolev space $\mathcal{W}^{p,2}([0,1])$, which satisfies (A5) for any $p \geq 1$. If $f_0 \in \mathcal{W}^{p,2}([0,1])$, under the assumptions of \citet{Zhao:2012} we find $||\widetilde{f}_K - f_0|| = O(K^{-p})$ and for the $\ell_1$-penalty on the wavelet coefficients we have $\lambda \mathcal{J}(\widetilde{f}_K) = \lambda^{1-r/2}$ for some $r \in (0,2)$ leading to
\begin{align*}
|\pi(\widehat{f}_n, f_0)|^2 = O_{\Pr}\left( \frac{K \log n}{n} \right) + O_{\Pr}\left(\frac{1}{K^{2p}}\right) + O_{\Pr}\left(\lambda^{1-r/2}\right).
\end{align*}
The regularization bias is now different from the two previous examples because of the thresholded wavelet coefficients, see \citet{Zhao:2012} for more details.

\subsection{Generalization to higher dimensions}

For clarity, we have so far focused on the case of $\mathcal{I} = [0,1]$, that is, a stochastic process $X:[0,1] \times \Omega \to \mathbb{R}$. However, our estimation method and theoretical development permit important extensions to random fields. In particular, let $\mathcal{I}$ now denote a subset of $\mathbb{R}^d$ with $d \geq 1$ and consider the multidimensional extension of \eqref{eq:1} given by
\begin{align*}
Y = \alpha_0 + \int_{\mathcal{I}} X(\mathbf{t}) f_0(\mathbf{t}) d\mathbf{t} + \sigma_0 \epsilon,
\end{align*}
for some $f_0 \in \mathcal{L}^2(\mathcal{I})$. 

For example, for $\mathcal{I}=[0,1]^d$ an approximating subspace may be easily constructed by taking tensor products of univariate approximating subspaces, that is, we consider the subspace $\Theta_K = \Theta_{K_1} \otimes \ldots  \otimes \Theta_{K_d}$. A multivariate penalized MM-estimator may now be defined as follows 
\begin{align}
\label{eq:11}
(\widehat{\alpha}_n, \widehat{f}_n ) = \argmin_{\alpha \in \mathbb{R}, f \in \Theta_{\mathbf{K}} } \left[  \frac{1}{n} \sum_{i=1}^n \rho \left( \frac{Y_i - a - \langle X_i, f \rangle}{\widehat{\sigma}_n} \right)  + \mathcal{J}_{\boldsymbol{\lambda}}(f) \right],
\end{align}
where for $f,g\in \mathcal{L}^2(\mathcal{I})$, $\langle f, g \rangle = \int_{\mathcal{I}} f(\mathbf{t})g(\mathbf{t}) d\mathbf{t}$ and $\mathcal{J}_{\boldsymbol{\lambda}}: \Theta_\mathbf{K} \to \mathbb{R}_{+}$ is an appropriate penalty functional that depends on a vector of smoothing parameters $\boldsymbol{\lambda}=(\lambda_1, \ldots, \lambda_d)$.

Inspection of the proofs of Theorems~\ref{Thm:1} and \ref{Thm:2} above reveals that the these theorems carry over to multivariate MM-estimators without additional difficulty, under straightforward adaptations of assumptions (A4), (A5), (A6) and (A7). A uniform law of large numbers as in Lemma~\ref{Lem2} of the Appendix (with $\Theta_{\mathbf{K}}$ replacing $\Theta_K$) can then be derived, provided that $\prod_{j=1}^d K_j = o(n)$. Combined with the adapted assumptions, this law allows to show that Theorem~\ref{Thm:1} remains valid in the multivariate setting, i.e.,
$\sup_{\mathbf{x} \in \mathcal{I}}|\widehat{f}_n(\mathbf{x}) - f_0(\mathbf{x})| \xrightarrow{P} 0$. Furthermore, the bracketing integral of the class of functions of $(X,y) \in \mathcal{L}^2(\mathcal{I}) \times \mathbb{R}$ given by
\begin{align*}
\mathcal{G}_{n, c, \delta} = \left\{\rho\left( \frac{y-\langle X, f \rangle}{\sigma} \right) - \rho\left( \frac{y-\langle X, \widetilde{f}_\mathbf{K} \rangle}{\sigma} \right),\ f \in \Theta_\mathbf{K}, \ ||f-\widetilde{f}_\mathbf{K}|| \leq c,\ |\sigma-\sigma_0| \leq \delta \right\}
\end{align*}
from $0$ to every small $\epsilon>0$ behaves like $C_0\epsilon \log^{1/2}\left( \frac{1}{\epsilon} \right) \prod_{j=1}^d K_j$ for some constant $C_0$. Therefore, the argument in the proof of Theorem~\ref{Thm:2} yields
\begin{align*}
|\pi(\widehat{f}_n, f_0)|^2 = O_{\Pr} \left( \frac{\log n \prod_{j=1}^d K_j }{n} \right) + O_{\Pr}(||\widetilde{f}_\mathbf{K}-f_0||
^2) + O_{\Pr}(\mathcal{J}_{\boldsymbol{\lambda}}(\widetilde{f}_\mathbf{K})).
\end{align*}
	The inflation of the variance term is a manifestation of the curse of dimensionality and translates into comparatively lower rates of convergence for large $d$. Appropriate roughness penalties on $\Theta_{\mathbf{K}}$ are thin-plate and tensor product penalties, for example \citep[see][Chapter 5 for more details]{Wood:2017}.

\section{A Monte-Carlo study}
\label{sec:4}

In our simulation scenarios we examine the effects of  the shape of the true coefficient function and outlying observations on four functional regression estimators. The first estimator we consider is the proposed penalized MM-estimator, denoted by $\widehat{f}_{PMM}$, based on a spline subspace with a difference (P-spline) penalty and settings described in Section~\ref{sec:2}. We compare this estimator to the unpenalized B-spline estimator of \citet{Boente:2020}, the robust reproducing kernel estimator of \citet{Shin:2016} and the FPCR estimator of \citet{reiss2007functional}. We now briefly review these three estimators.

Consider the spline subspace $\Theta_{K+p}$ given in Example 1 above. Then, the MM-estimator proposed by \citet{Boente:2020} solves
\begin{align*}
(\widehat{\alpha}_{MM}, \widehat{f}_{MM} ) = \argmin_{\alpha \in \mathbb{R}, f \in \Theta_{K+p} } \left[  \frac{1}{n} \sum_{i=1}^n \rho \left( \frac{Y_i - a - \langle X_i, f \rangle}{\widehat{\sigma}_n} \right)  \right].
\end{align*}
Here, $\rho$ is a bounded loss function, the Tukey bisquare in our implementation, and $\widehat{\sigma}_n$ is the S-scale estimate discussed in Section~\ref{sec:2}. The dimension of the subspace, which is proportional to the number of interior knots, acts as the tuning parameter. As proposed by \citet{Boente:2020}, the  interior knots are spread within $[0,1]$ in an equidistant manner and the number of interior knots is selected according to a BIC-type criterion. 

A robust penalized estimator according to the smoothing-spline principle was proposed by \citet{Shin:2016} as a robustification of the corresponding least-squares estimator of \citet{Yuan:2010}. This estimator solves 
\begin{align*}
(\widehat{\alpha}_{SMSP}, \widehat{f}_{SMSP} ) = \argmin_{\alpha \in \mathbb{R}, f \in \mathcal{W}^{2,2}([0,1]) } \left[  \frac{1}{n} \sum_{i=1}^n \rho \left( \frac{Y_i - a - \langle X_i, f \rangle}{\widehat{\sigma}_n} \right)  + \lambda ||f^{\prime \prime}||^2 \right],
\end{align*}
where $\mathcal{W}^{2,2}([0,1])$ is a Sobolev space of functions with squared integrable second derivative, $\rho$ is the bisquare loss function and $\widehat{\sigma}_n$ is the MAD obtained from an initial Huber smoothing spline fit to the data \citep[see][for details]{Shin:2016}. It can be shown that the solution is of the form
\begin{align*}
f(t) = d_1 + d_2 t + \sum_{i=1}^n c_i \int_{0}^1 X_i(s) K(t,s) ds,
\end{align*}
for some $d_1,d_2 \in \mathbb{R}$ and $(c_1,\dots,c_n) \in \mathbb{R}^n$, with $K : \mathbb{R} \times \mathbb{R} \to \mathbb{R}$ the reproducing kernel of $\mathcal{W}^{2,2}([0, 1])$ \citep[see][for more details]{Hsing:2015}. The regularization parameter $\lambda$ is selected via generalized cross-validation. For all three estimators $\widehat{f}_{PMM}$, $\widehat{f}_{MM}$ and $\widehat{f}_{SMSP}$, the tuning constant in the Tukey-bisquare loss function was set equal to 4.685, corresponding to $95\%$ efficiency in the location model under Gaussian errors.

Let $t_1, \ldots, t_k \in [0,1]$ denote the points of discretization of the curves $\left\{X_i(t) \right\}_{i=1}^n$ and let $\mathbf{S} := \left[X_i(t_j) \right]_{i, j} $ denote the $n \times k$ matrix of discretized signals. Furthermore, let $\mathbf{B}$ denote a $k \times (K+4)$ matrix of cubic B-spline functions evaluated at the discretization points and let $\mathbf{B}(x) = ( B_1(x), \ldots, B_{K+4}(x) )^{\top} $ be the vector of cubic B-spline functions evaluated at $x$. 
Let $\mathbf{V}_A$ be the matrix containing the first $A$ right singular vectors of $\mathbf{S} \mathbf{B}$,
then the least-squares based FPCR estimator minimizes
\begin{equation*}
\left\|\mathbf{Y}-\alpha \mathbf{1} - \mathbf{S} \mathbf{B} \mathbf{V}_A \boldsymbol{f}_1 \right\|^2_E + \lambda \ \boldsymbol{f}^{\top}_1 \mathbf{V}_A^{\top} \mathbf{D}_2 \mathbf{V}_A \boldsymbol{f}_1,
\end{equation*}
with respect to $\boldsymbol{f}_1$.
 The estimator for the coefficient function is then given by $\widehat{f}_{FPCR}(t) = \mathbf{B}^{\top} (t)\widehat{\boldsymbol{V}}_A \widehat{\boldsymbol{f}}_1$. Free selection of the smoothing parameters $K$, $A$ and $\lambda$ is computationally intensive. Hence, in practice the procedure is implemented by fixing $K= 36$, selecting $A$ such that the explained variation of $\mathbf{S}\mathbf{B}$ is 0.99 and estimating $\lambda$ by restricted maximum likelihood.

In order to compare the four competing estimators we have generated curves according to the truncated Karhunen-Loève decomposition given by
\begin{align}
X(t) = \sum_{j=1}^{50} j^{-1} Z_j \sqrt{2} \cos((j-1)\pi t),
\label{curve_design}
\end{align}
where the $Z_j$ are random variables whose distribution is varied according to the scenarios outlined below. These curves are combined with each of the following four coefficient functions
\begin{enumerate}
\item $f_1(t) = \sin(2 \pi t)$
\item $f_2(t) = t^2 \phi(t, 0, 1) $
\item $f_3(t) = 1/(1+e^{-20(t-0.5)})$
\item $f_4(t) = -\phi(t, 0.2, 0.03) + 3 \phi(t, 0.5, 0.03) + \phi(t, 0.75, 0.04)$,
\end{enumerate}
where $\phi(t, \mu, \sigma)$ denotes the Gaussian density with mean $\mu$ and standard deviation $\sigma$. These regression functions represent a variety of different characteristics: $f_1$ is a sinusoid, $f_2$ is almost a straight line with some curvature near the boundaries, $f_3$ is a sigmoid and $f_4$ is bumpy. Due to its local characteristics, $f_4$ is much more difficult to estimate precisely than the other functions. 

\begin{figure}[H]
\centering
\subfloat{\includegraphics[width = 0.495\textwidth]{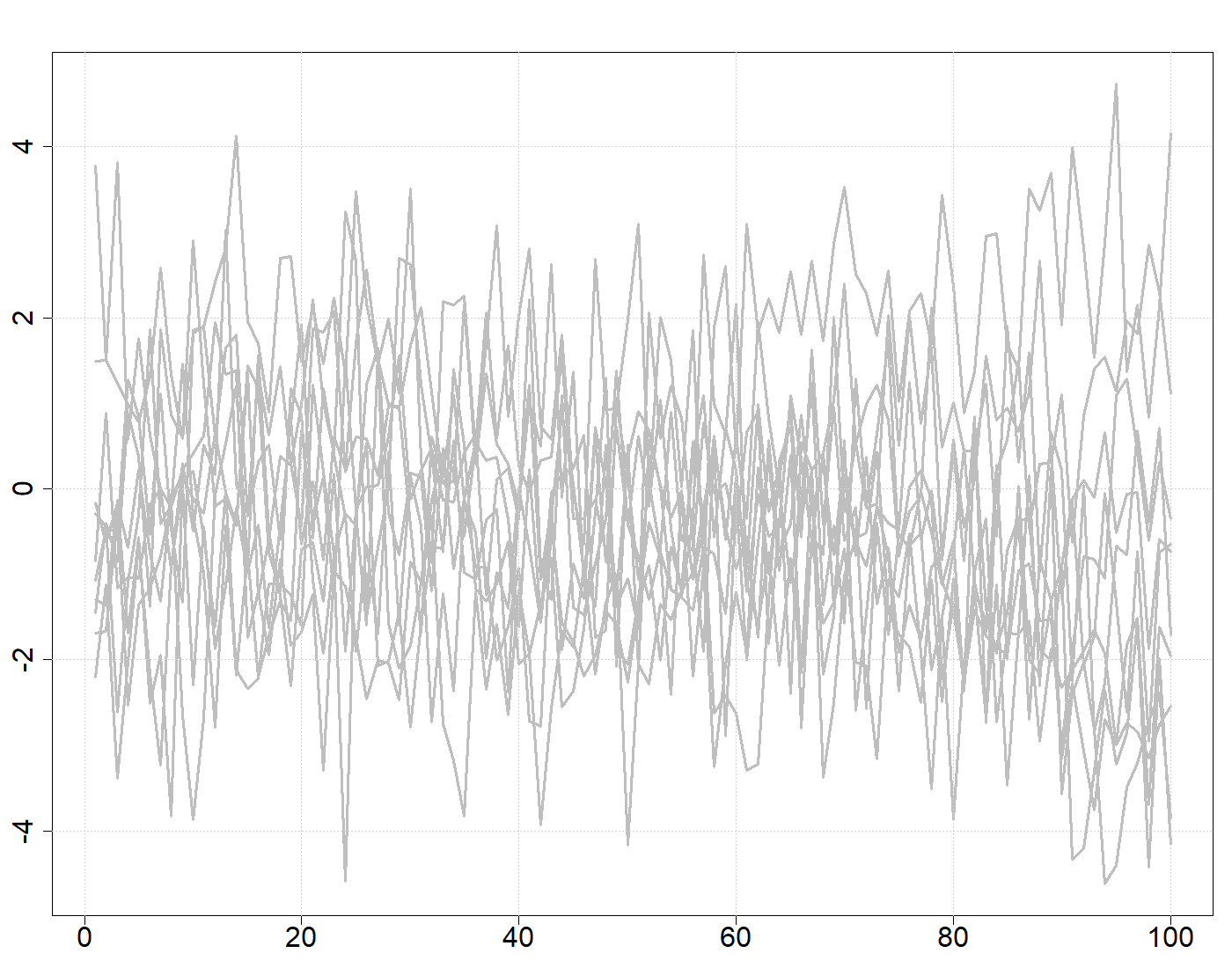}} \ 
\subfloat{\includegraphics[width = 0.495\textwidth]{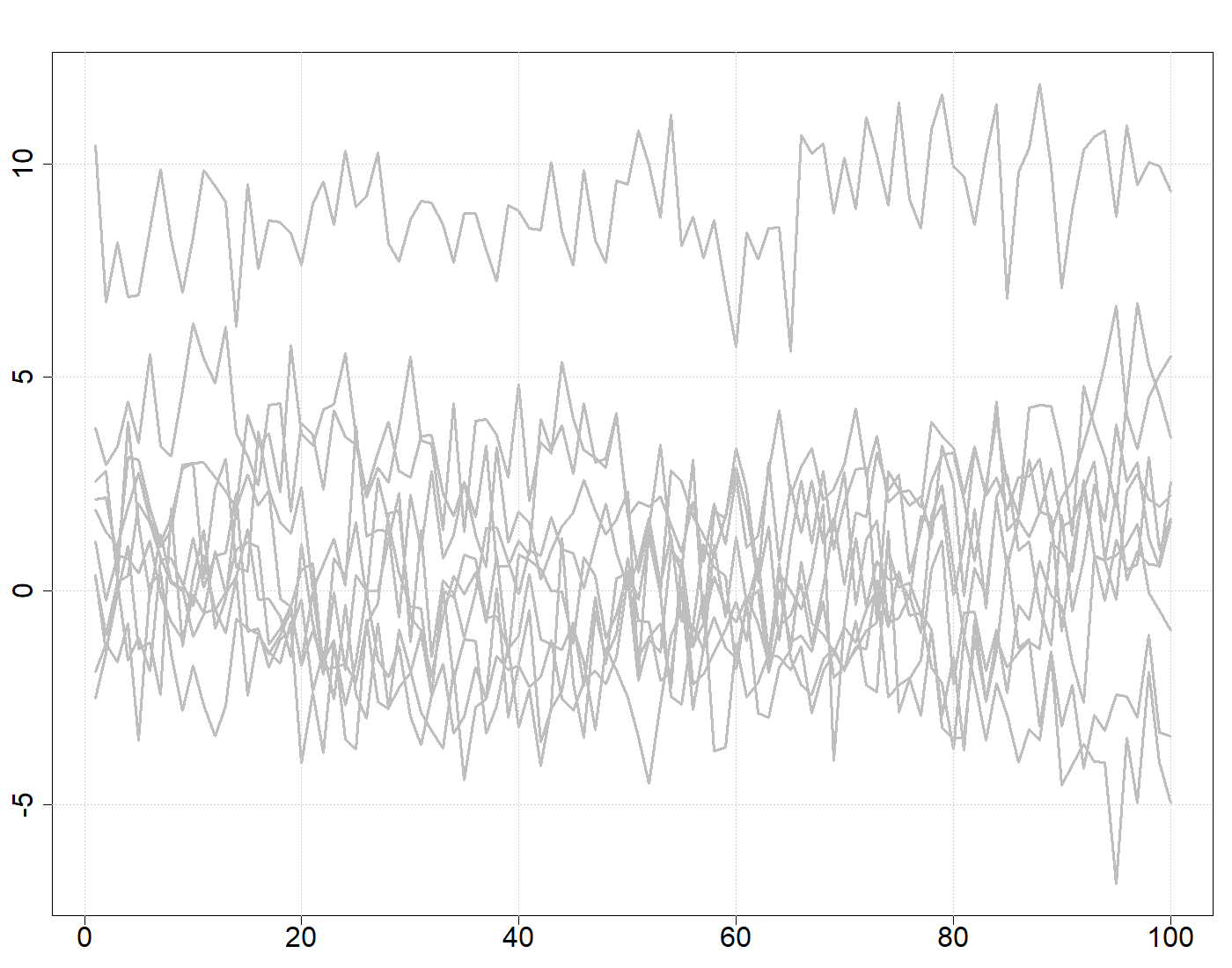}}
\caption{Two representative samples of curves with $Z_j \sim \mathcal{N}(0,1)$ and $Z_j \sim t_{5}$ on the left and right respectively.}
\label{fig:1}
\end{figure}

We set $\sigma_0=1$ in \eqref{eq:1} and consider the following scenarios for the scores $Z_j$ in (\ref{curve_design}) and the errors $\epsilon_i$ in \eqref{eq:1}:
\begin{itemize}
\item[Scen. 1] The $Z_j$ and $\epsilon_i$ both follow standard Gaussian distributions.
\item[Scen. 2] The $Z_j$ follow a standard Gaussian distribution and the $\epsilon_i$ follow a $t_3$-distribution.
\item[Scen. 3] The $Z_j$ follow standard Gaussian distributions and the $\epsilon_i$ follow a Gaussian mixture distribution with density $0.9 \phi(t, 0, 1) + 0.1 \phi(t, 14, 1)$.
\item[Scen. 4] The $Z_j$ follow a $t_{5}$-distribution and the $\epsilon_i$ follow the same Gaussian mixture distribution as in the previous scenario.
\end{itemize}
These scenarios reflect increasingly severe contamination. The first scenario portrays the ideal situation of light-tailed predictors and error, the second scenario introduces mild vertical outliers and the third scenario yields more severe contamination. Lastly, by perturbing the distribution of the $Z_j$, the fourth scenario combines vertical outliers and leverage points. For a better appreciation of  the effect of the distribution of the $Z_j$ on the shape of the curves, figure~\ref{fig:1} plots two representative samples of curves with the $Z_j$ following Gaussian and $t_5$ distributions.

To handle the curves practically we have discretized them in 100 equidistant points $t_1, \ldots, t_{100}$, within the $[0,1]$-interval and computed all related inner products using  Riemann approximations. To evaluate the performance of the estimators we calculate their mean-square error (MSE) given by
\begin{equation*}
\MSE = 100^{-1} \sum_{j=1}^{100} | \widehat{f}(t_j) - f_0(t_j) |^2.
\end{equation*}
This statistic is an approximation to the $\mathcal{L}^2([0,1])$-distance $||\widehat{f}-f_0||^2$. Table~\ref{tab:1} below presents average and median MSEs for all of our settings for $n=150$ and $1000$ replications.

There are several interesting conclusions that emerge from this study. First, the performance of the least-squares based $\widehat{f}_{FPCR}$ heavily depends on the distribution of the scores and errors. The estimator performs best under light-tailed distributions, but quickly loses ground when faced with slightly heavier tails, e.g., with a $t_3$-distribution in the errors. This performance is in line with expectations regarding least-squares estimators which are known to be very sensitive to even a small number of mildly outlying observations. By contrast, the robust estimators $\widehat{f}_{MM}$ and $\widehat{f}_{PMM}$ maintain a much more steady performance. In particular, note that $\widehat{f}_{PMM}$ matches the performance of $\widehat{f}_{FPCR}$ in Gaussian data and exhibits a high degree of resistance towards atypical data in all settings considered.

\begin{table}[H]
\centering
\resizebox{\columnwidth}{!}{%
\begin{tabular}{c c c c c c c c c c c }
\multicolumn{1}{c}{}  & &  \multicolumn{2}{c}{$\widehat{f}_{FPCR}$} & \multicolumn{2}{c}{$\widehat{f}_{RKHS}$} & \multicolumn{2}{c}{$\widehat{f}_{MM}$} & \multicolumn{2}{c}{$\widehat{f}_{PMM}$}  \\ \\[-2ex]
 & &  Mean & Median & Mean & Median & Mean & Median & Mean & Median   \\ \\ [-1.2ex]
\multirow{3}{*}{$f_1$} & Scen. 1  & 0.257 &  0.236 & 2.020 & 1.735 &   0.145 & \textbf{0.125} & 0.288 &  0.281 \\
& Scen. 2 & 0.508  & 0.446 & 2.083 & 1.790 & 0.186 & \textbf{0.156} &  0.336 & 0.322   \\
& Scen. 3  & 1.746 & 1.506 & 2.083 & 1.789 & 0.160 & \textbf{0.135} & 0.294  & 0.287 \\
& Scen. 4  & 1.366 &  1.226 & 2.672 & 2.520 & 0.127 & \textbf{0.112} &  0.257 & 0.252
 \\  \\[-1.2ex]
\multirow{3}{*}{$f_2$} & Scen. 1  & 0.112 &  0.097 & 0.027 & 0.016 &  0.027 & \textbf{0.015} &  0.026 & 0.019 \\
& Scen. 2 & 0.217  & 0.187 & 0.126 & 0.103 & 0.046 & 0.026 &  0.033 & \textbf{0.025}   \\
& Scen. 3  & 0.731 & 0.641 &  0.092 & 0.080 & 0.034 &  0.021 &  0.026 & \textbf{0.020} \\
& Scen. 4  & 0.653 & 0.584 & 0.061 & 0.051 & 0.024 & 0.017 &  0.019 & \textbf{0.014}
 \\ \\[-1.2ex]
\multirow{3}{*}{$f_3$} & Scen. 1  & 0.306 &  0.274 & 0.450 &  0.428 &  0.258 &  0.229 & 0.189 & \textbf{0.173} \\
& Scen. 2 &  0.628 & 0.548 &0.489 & 0.465 & 0.374 & 0.294 & 0.264 & \textbf{0.237} \\
& Scen. 3  &  1.887 & 1.700 & 0.507 & 0.475 & 0.230 & 0.201 &  0.147 & \textbf{0.138} \\
& Scen. 4  &  1.887 & 1.700 & 0.507 & 0.475 & 0.230 & 0.201 &  0.147 & \textbf{0.138} \\ \\[-1.2ex]
\multirow{3}{*}{$f_4$} & Scen. 1  & 10.85 &   \textbf{9.319} & 42472 &40820 &  8222 & 8189 & 10.00 & 9.881 \\
& Scen. 2 &  15.29  & 13.04 & 42748 & 41026 & 8244 & 8196 &  11.81 & \textbf{11.551}   \\
& Scen. 3  & 46.84 & 45.67 &  42634 & 40940 & 8240 & 8208 &  10.05 & \textbf{9.862} \\
& Scen. 4  &  43.16 & 41.83 & 45518 & 42318 & 8197 & 8156 &   9.606 & \textbf{9.470}
\\ 
\end{tabular}}
\caption{Mean and median of the MSE ($\times 1000)$ for the competing estimators over 1000 datasets of size $n=150$. Best median performances are in bold.}
\label{tab:1}
\end{table}

Comparing the performance of the robust estimators in more detail reveals that $\widehat{f}_{MM}$ and $\widehat{f}_{PMM}$ outperform $\widehat{f}_{RKHS}$ in almost all settings considered. Moreover, it can be seen that the simpler $\widehat{f}_{MM}$ has an edge over $\widehat{f}_{PMM}$ whenever the regression function is relatively simple, that is, without any local characteristics. In particular, for coefficient function $f_1$, the estimator $\widehat{f}_{MM}$ performs almost twice as well as $\widehat{f}_{PMM}$. The reason for this difference is that for such simple situations the regression function can be approximated with only a handful of equidistant knots and this gives $\widehat{f}_{MM}$ an advantage over $\widehat{f}_{PMM}$. However, this strategy can go very awry when one is confronted with a more complex coefficient function such as the bumpy coefficient function $f_4$, for instance. In such cases, $\widehat{f}_{MM}$ becomes completely unreliable and gets outperformed by $\widehat{f}_{PMM}$ by a wide margin.

\begin{figure}[H]
\centering
\subfloat{\includegraphics[scale = 0.3, width = 0.49\textwidth]{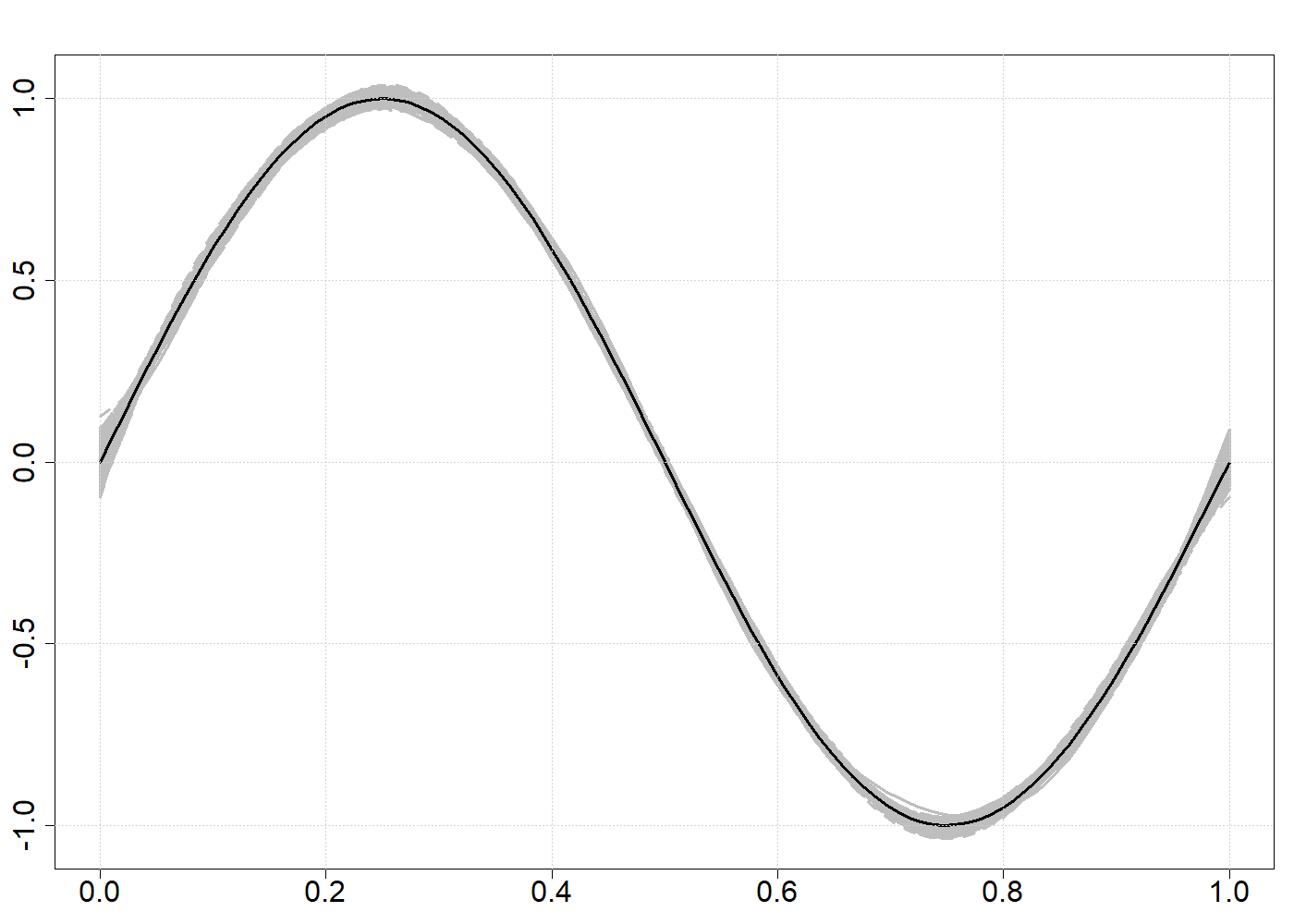}} 
\subfloat{\includegraphics[scale = 0.3, width = 0.49\textwidth]{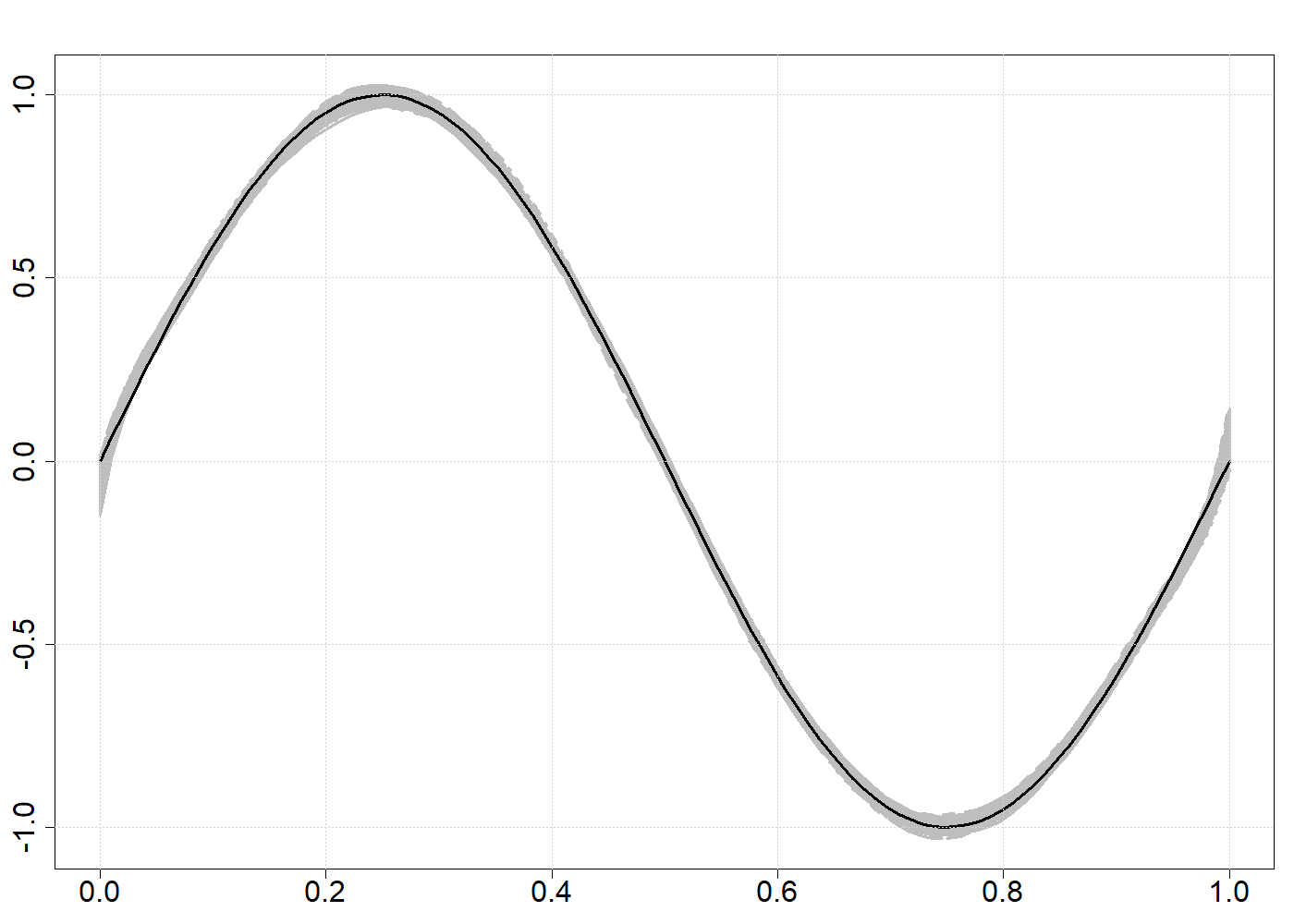}} \\		
\subfloat{\includegraphics[scale = 0.3, width = 0.49\textwidth]{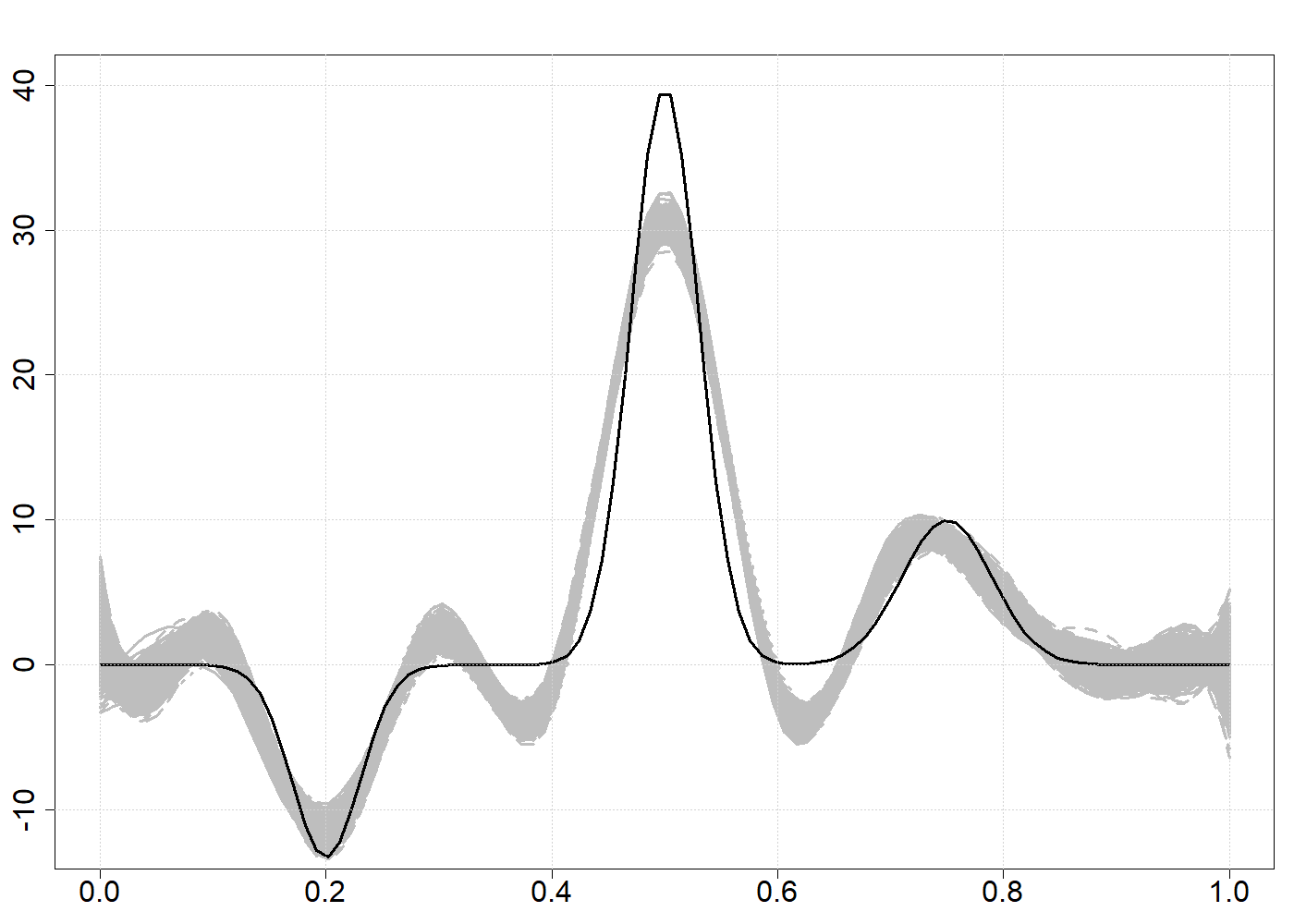}}  
\subfloat{\includegraphics[scale = 0.3, width = 0.49\textwidth]{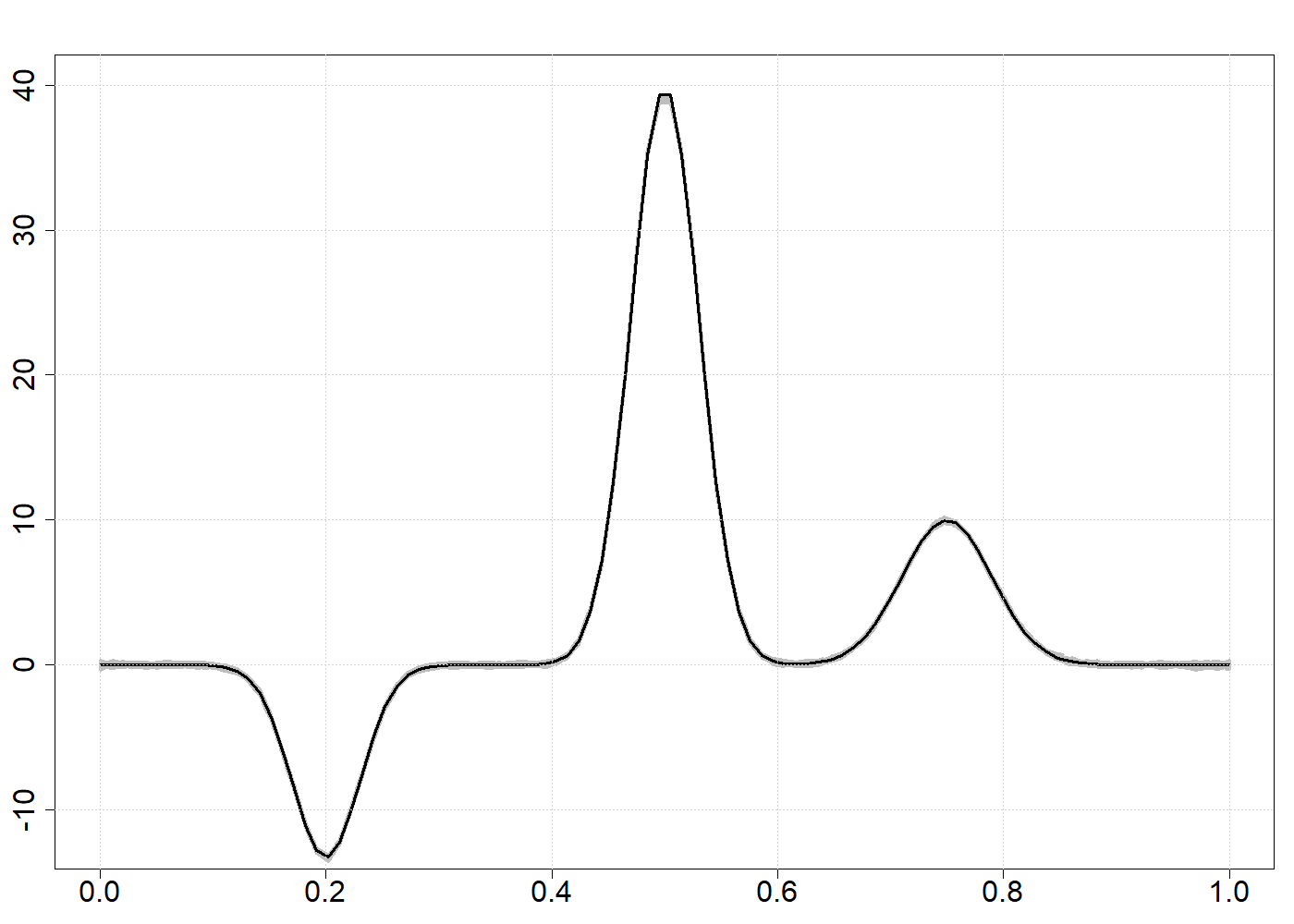}} 
\caption{$\widehat{f}_{MM}$ (left) and $\widehat{f}_{PMM}$ (right) estimates for $f_1$ (top) and $f_4$ (bottom) under Scenario 1. The solid black line in each plot corresponds to the true coefficient function.}
\label{fig:2}
\end{figure}

To illustrate the key differences between $\widehat{f}_{MM}$ and $\widehat{f}_{PMM}$, Figure~\ref{fig:2} presents the 1000 estimates for $f_1$ and $f_4$ obtained under Scenario 1 along with the true coefficient functions. It is worth noticing how minor the differences are in the case of $f_1$ and how variable the $\widehat{f}_{MM}$-estimates are in the case of $f_4$.  These estimates tend to lack the right amount of smoothness, thereby missing the local characteristics of $f_4$. This in turn leads to the poor performance seen in Table~\ref{tab:1}. The performance of $\widehat{f}_{MM}$ in that case could be improved by a more careful selection of the location of the knots, but this would inevitably lead to a much increased computational burden. Overall, these simulation results indicate that $\widehat{f}_{PMM}$ is a viable alternative to $\widehat{f}_{FPCR}$ in regular data and remains reliable in a wider range of contaminated data settings than its unpenalized alternative $\widehat{f}_{MM}$.

\section{Real data example: archaeological glass vessels}
\label{sec:5}

In this section we apply the proposed penalized  estimator to the popular glass dataset. This dataset contains measurements for 180 archaeological glass vessels (15th to 17th century) that were recently excavated from the old city of Antwerp, which prior to the tumultuous 17th century was one of the largest ports in Europe with extensive ties to commercial centres all over the continent, see \citep{Janssens:1998} for more background. The dataset is freely available in \textsf{R}-packages \texttt{chemometrics} \citep{Filzmoser:2017} and \texttt{cellWise} \citep{C:2019}.

\begin{figure}[H]
\centering
\subfloat{\includegraphics[scale = 0.3, width = 0.49\textwidth]{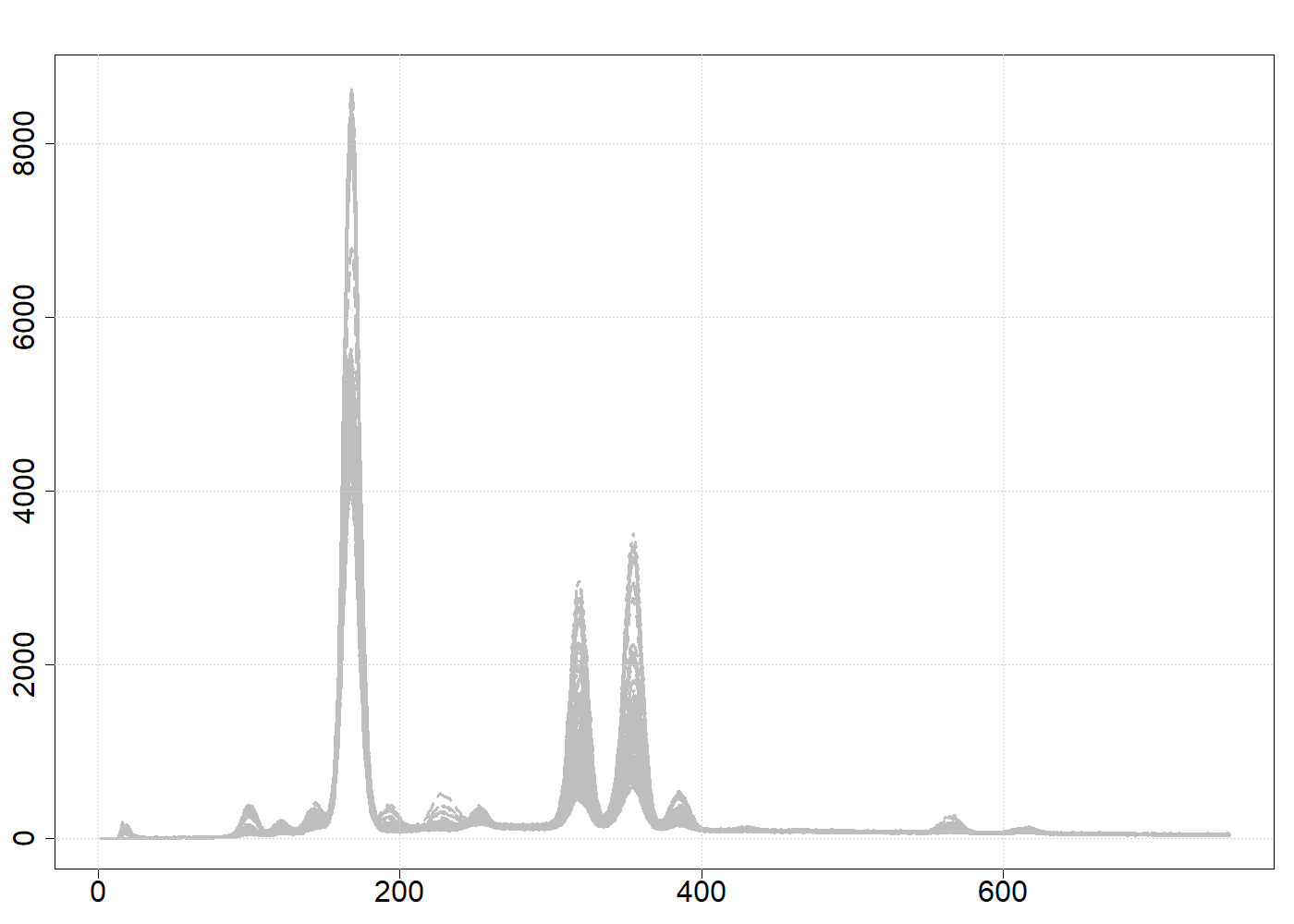}} \
\subfloat{\includegraphics[scale = 0.3, width = 0.49\textwidth]{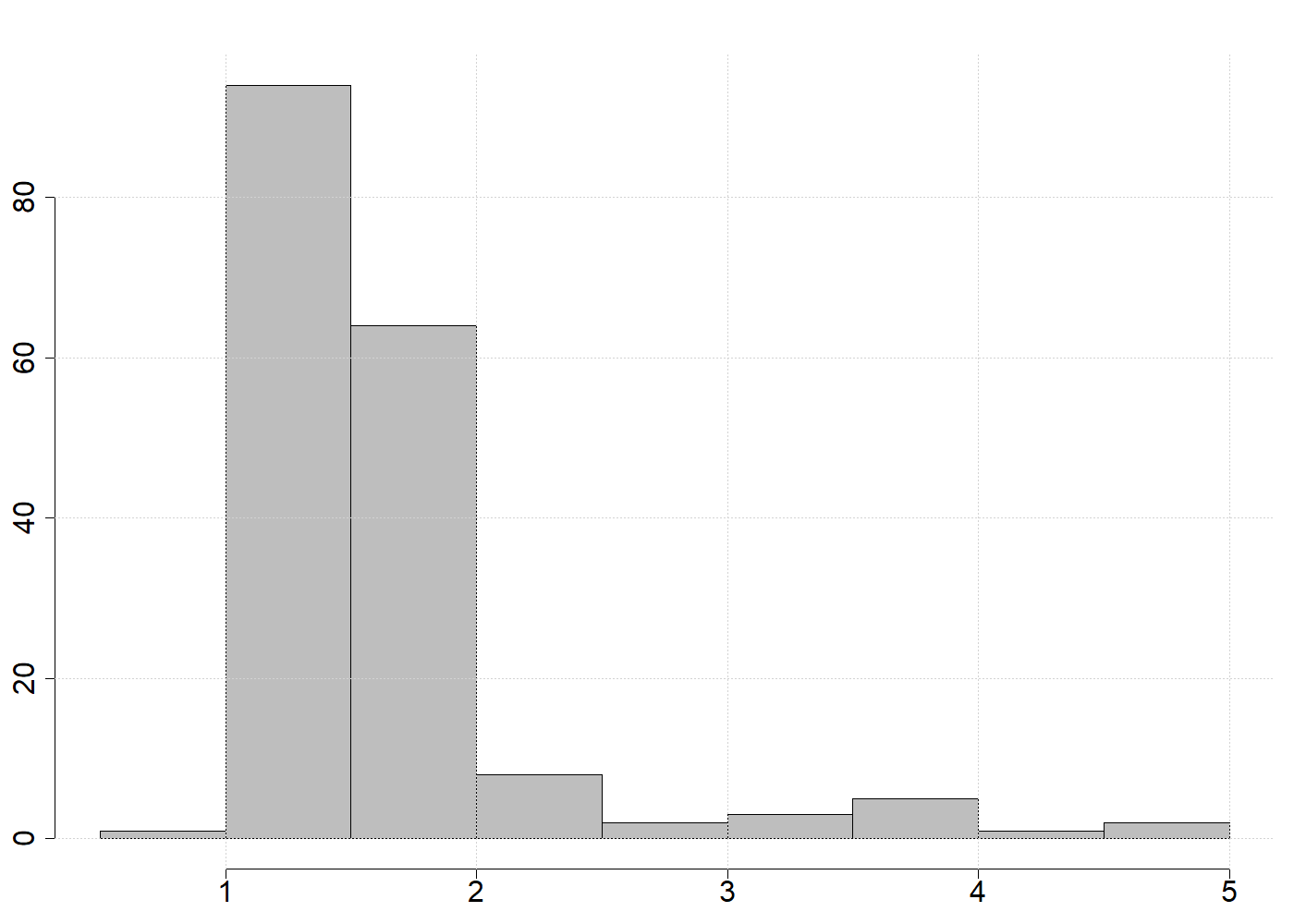}} \\		
\subfloat{\includegraphics[scale = 0.3, width = 0.49\textwidth]{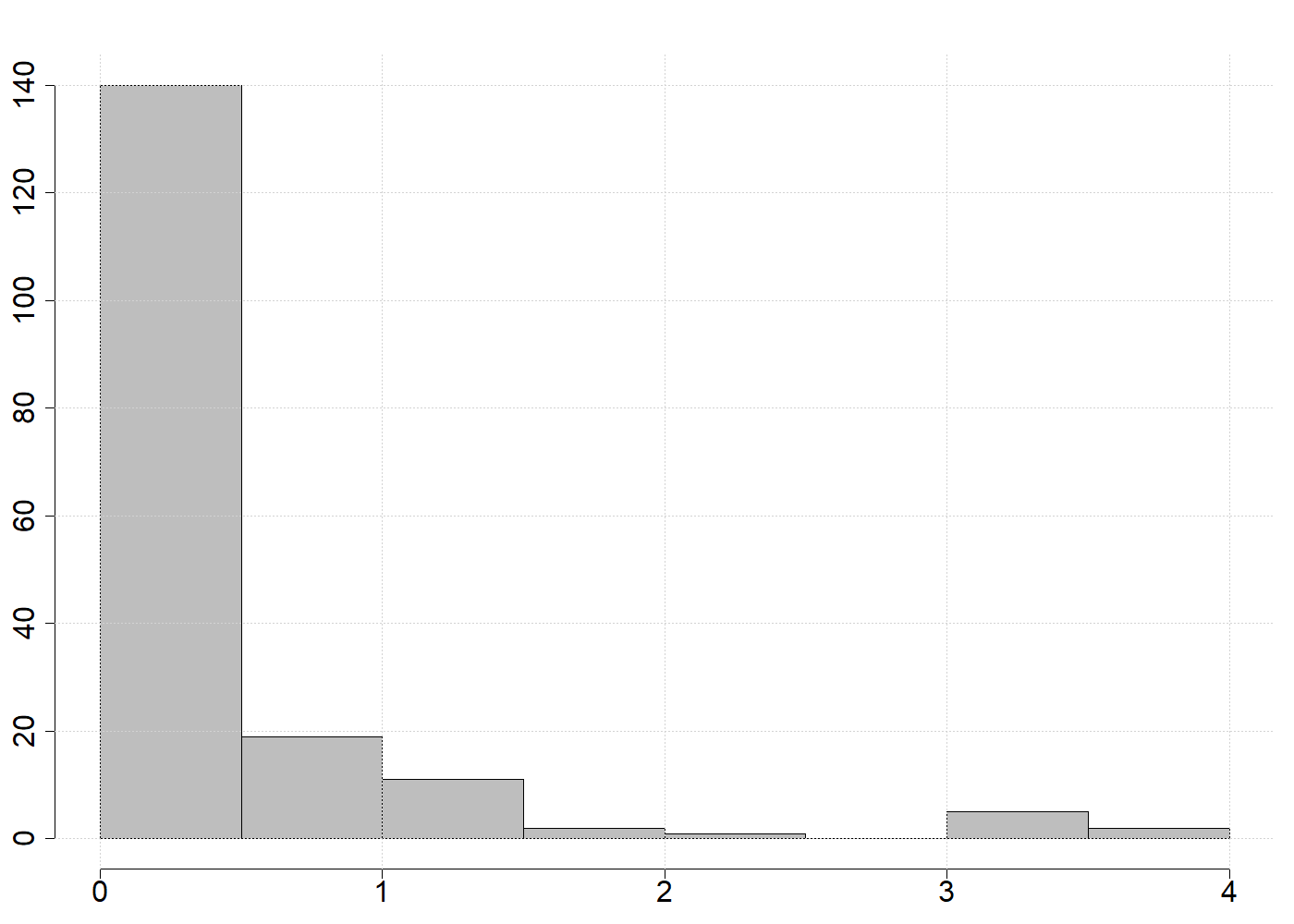}} \ 
\subfloat{\includegraphics[scale = 0.3, width = 0.49\textwidth]{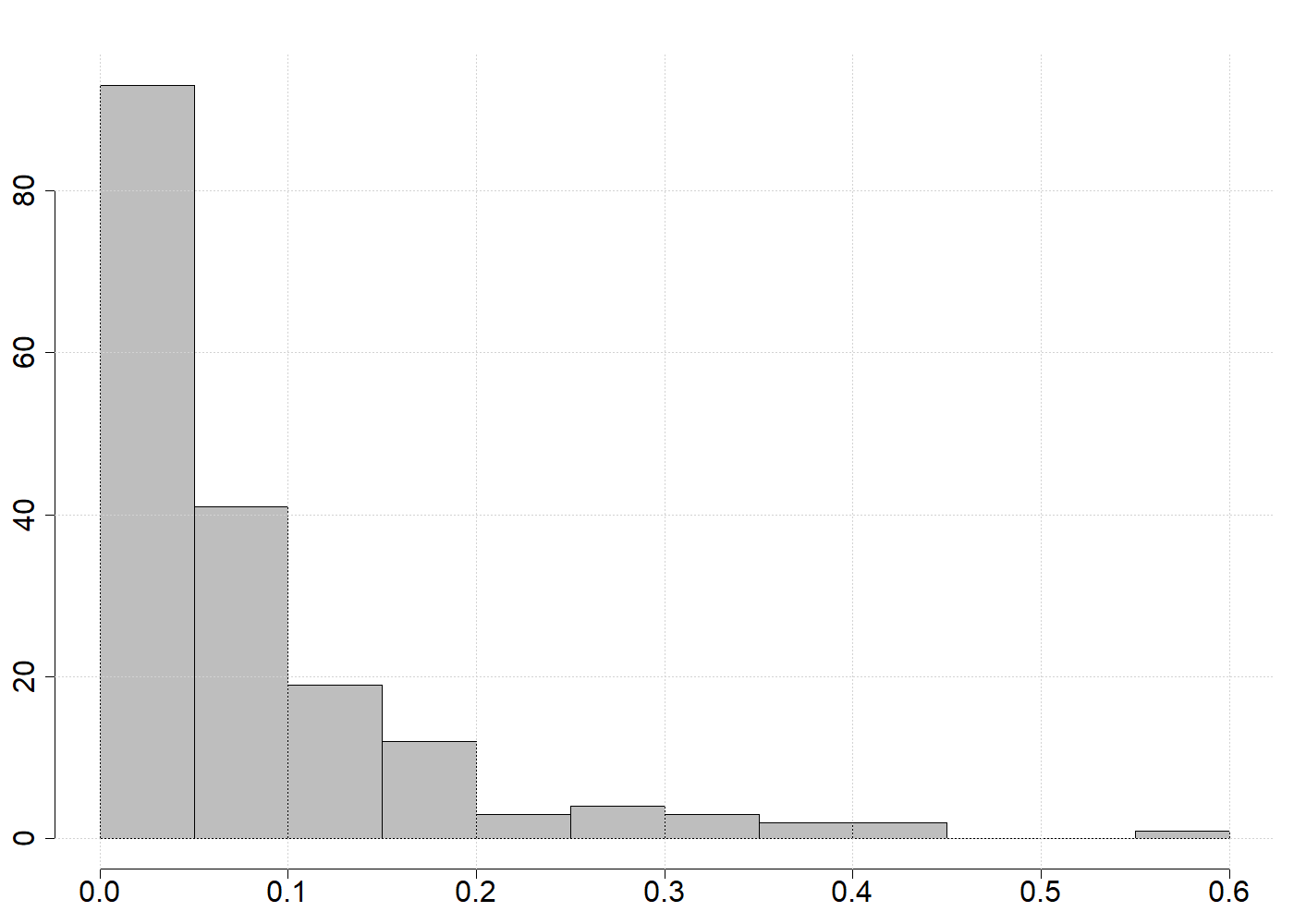}} 
\caption{Glass spectra and 3 chemical compounds (Al2O3, P2O5, BaO).}
\label{fig:3}
\end{figure}

For each of the vessels we are in possession of near-infrared spectra with 750 wavelengths, along with the values of 13 chemical compounds which are crucial for the determination of the type of glass as well as its origin. A reduced form of this dataset with only the non-null spectra was analyzed by \citet{Maronna:2013}. However, here we avoid any preprocessing of the data. Plots of the spectra and some of the chemical compounds are given in Figure \ref{fig:3}. By examining the heights of the peaks in the spectra it may be conjectured that there are three types of glass in the sample, which is indeed the key finding of \citet{Janssens:1998}. The histograms of the chemical compounds further indicate that the distributions of these responses are right-skewed with several potential outliers.  

\begin{table}[H]
\centering
\resizebox{\columnwidth}{!}{%
\begin{tabular}{ccc ccc ccc ccc cc }
& Na2O & MgO & Al2O3 & SiO2 & P2O5 & SO3 & Cl & K2O & CaO & MnO &  Fe2O3 & BaO & PbO \\ \\[-1.4ex]
$\widehat{f}_{PMM}$ &  \textbf{0.531} & 0.253 & \textbf{0.065} & \textbf{0.436} & \textbf{0.043} & 0.043 &  \textbf{0.013} & \textbf{0.133} & \textbf{0.168} &  \textbf{0.018} & \textbf{0.015} & \textbf{0.017} & \textbf{0.086} \\ \\[-1.4ex]
$\widehat{f}_{FPCR}$ & 0.768 & \textbf{0.170} & 0.086 & 0.513 & 0.054 & \textbf{0.040} &  0.016 & 0.158 & 0.187 & 0.021 & 0.027 & 0.029 & 0.097
\end{tabular}}
\caption{RMSPE(0.9) for the $13$ compounds. Best performances in bold.}
\label{tab:2}
\end{table}

We compare the predictive performance of $\widehat{f}_{PMM}$ and $\widehat{f}_{FPCR}$ for each of the 13 responses in this dataset. The results for the $\widehat{f}_{MM}$ and $\widehat{f}_{SMSP}$ estimators are not reported because they performed significantly worse than $\widehat{f}_{PMM}$ on this complex dataset.
To measure the prediction performance of the methods, we apply 5-fold cross-validation. 
For each chemical compound we then compute the $10\%$ trimmed root mean squared error of the predictions, denoted by RMSPE(0.9).  This trimming is essential to measure prediction performance of the regular data because some of the left-out observations can be outliers \citep{Khan:2010}. 

\begin{figure}[H]
\centering
\resizebox{\columnwidth}{!}{%
\includegraphics[ width = 0.495\textwidth]{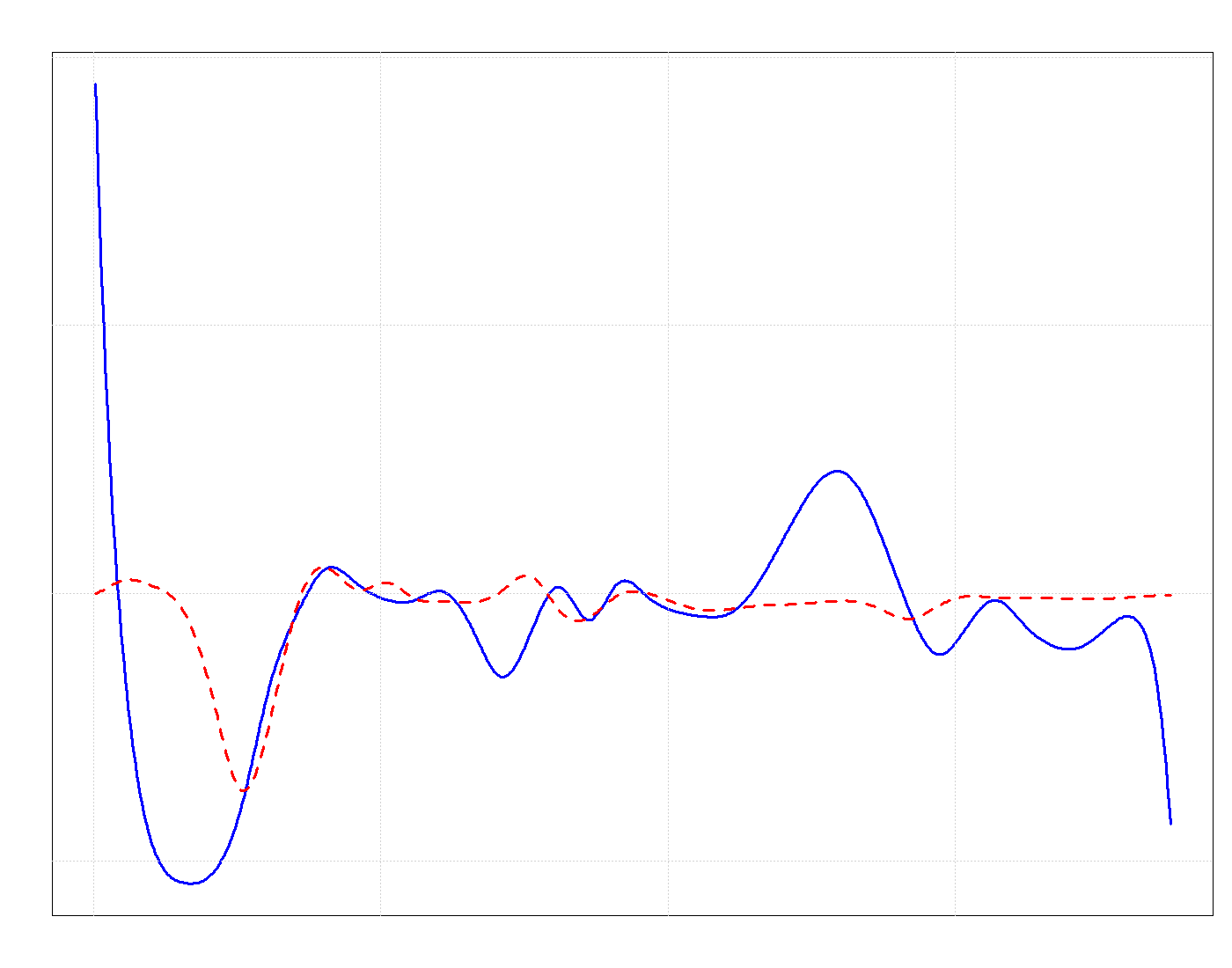}	
\includegraphics[ width = 0.495\textwidth]{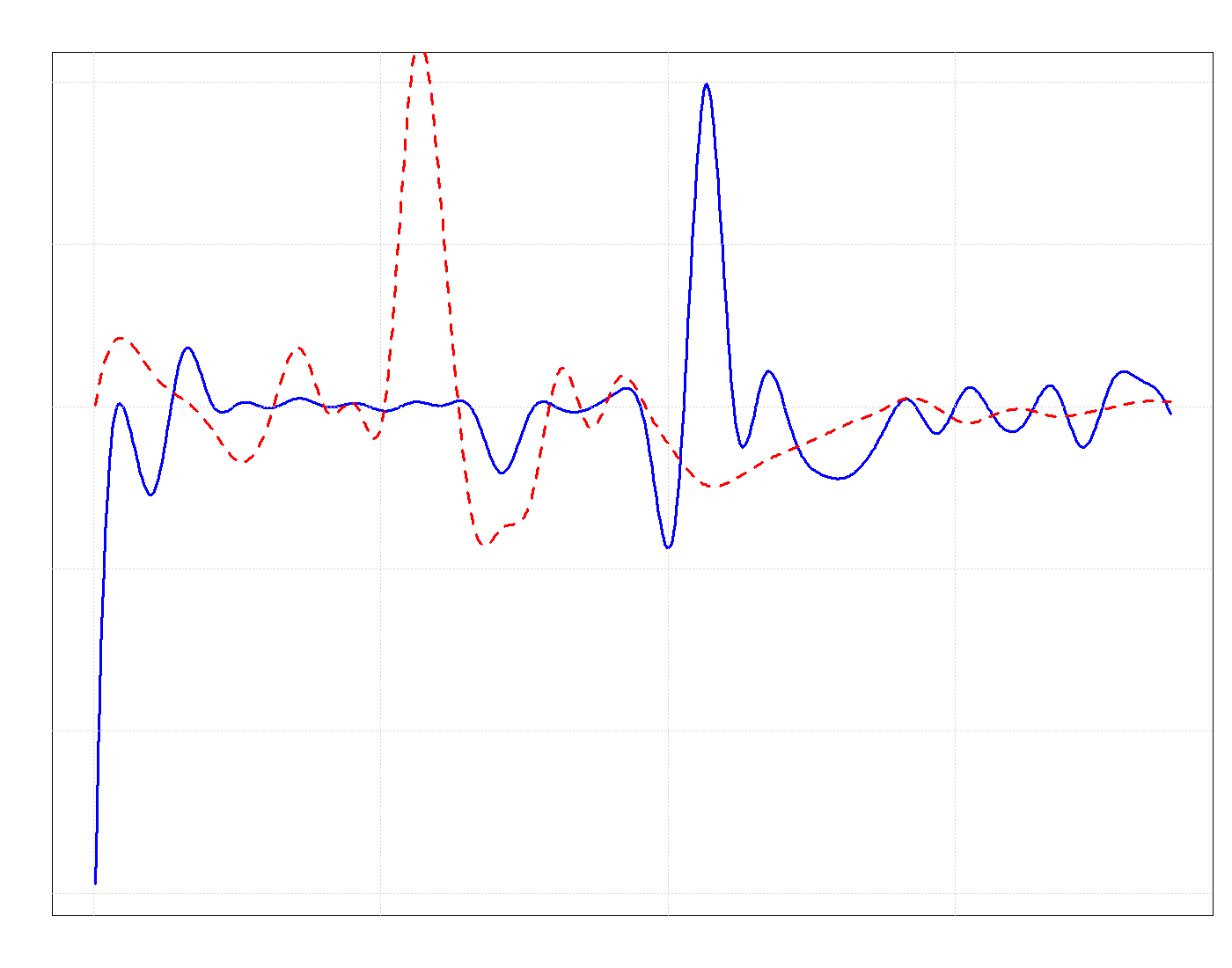}}
\caption{Functional regression estimates for P2O5 and BaO. The lines (\full, \denselydashed) correspond to $\widehat{f}_{PMM}$ and $\widehat{f}_{FPCR}$ estimates, respectively }
\label{fig:4}
\end{figure}

The results are summarized in Table \ref{tab:2}. It can be seen that $\widehat{f}_{PMM}$ outperforms $\widehat{f}_{FPCR}$ in all but two of the compounds. In practice, this means that in these cases $\widehat{f}_{PMM}$ provides better fits for the majority of the observations, thereby leading to better overall predictions of the observations following the model. The estimates of the coefficient functions for two of the chemical compounds are shown in Figure \ref{fig:4}. It is interesting to see that $\widehat{f}_{FPCR}$ produces smoother estimates than $\widehat{f}_{PMM}$, but this does not translate into better predictive performance. This suggests that the outlying observations lead to oversmoothed $\widehat{f}_{FPCR}$ estimates.

\section{Concluding remarks}
\label{sec:6}

We have shown that lower-rank penalized estimators based on bounded loss functions possess good theoretical properties, are computationally efficient and are capable of handling a diversity of complex problems, such as  
estimation of coefficient functions with local characteristics based on data with atypical observations.  Moreover, these important properties almost seamlessly extend to the case of scalar-on-function regression with random functions defined on multidimensional sets, such as images.

In future work we aim to further relax the assumptions underpinning the present theoretical development to take into account the often discrete sampling of functional data. This often neglected aspect of functional data has important practical and theoretical consequences, particularly when the discretization grid is small, see, e.g., \citet{Kalogridis:2021b} for the case of location estimation. A robust lower rank penalized regression estimator in this setting would constitute an effective and computationally efficient alternative to the smoothing spline estimators of \citet{Crambes:2009} and \citet{Maronna:2013}.

\section{Appendix: Proofs of the theorems}

\begin{lemma}[Fisher consistency]
\label{Lem1}
Assume that assumptions (A1), (A3) and (A6) hold. Then, for any $\sigma>0$ and $f \in \mathcal{B}([0,1])$, $M(f_0, \sigma) < M(f, \sigma)$, where 
\begin{align*}
M(f, \sigma) = \mathbb{E}\left\{ \rho \left( \frac{Y - \langle X, f \rangle }{\sigma} \right) \right\}.
\end{align*}
\end{lemma}

\begin{proof}
The proof is an adaptation of the corresponding proofs in \citet{Yohai:1987} and \citet{Boente:2020}. First, Lemma 3.1 of \citet{Yohai:1985} in combination with (A3) shows that the function $g(\alpha) = \mathbb{E}\{ \rho(\epsilon \sigma_0/\sigma - \alpha) \}$ has a unique minimum at zero, viz, for any $\alpha \neq 0$
\begin{align}
\label{eq:A2}
\mathbb{E}\left\{ \rho\left(\epsilon \frac{\sigma_0}{\sigma} - \alpha\right) \right\} > \mathbb{E}\left\{ \rho\left(\epsilon \frac{\sigma_0}{\sigma}\right) \right\}.
\end{align}
Fix $f \in \mathcal{B}([0,1])$, set $\mathcal{A}_0 = \left\{X: \Phi(X) = \langle X, f - f_0 \rangle = 0 \right\}$ and $\alpha(X) = \Phi(X)/\sigma$. Then,  using the independence of $\epsilon$ and $X$, it is not difficult to show that
\begin{align*}
M(f, \sigma) = \mathbb{E}\left\{ \rho\left( \epsilon \frac{\sigma_0}{\sigma}   \right) \right\} \Pr(\mathcal{A}_0) + \mathbb{E} \left\{ \mathbb{E}\left\{ \rho\left( \epsilon \frac{\sigma_0}{\sigma}- \alpha(X) \right) \vert X \right\} \mathcal{I}_{\mathcal{A}_0^c}(X) \right\},
\end{align*}
where $\mathcal{I}_B(\cdot)$ denotes the indicator function for a set $B$. Similarly, by the independence of $\epsilon$ and $X$ and \eqref{eq:A2}, for any $X \notin \mathcal{A}_0$,
\begin{align*}
\mathbb{E}\left\{ \rho \left(\epsilon \frac{\sigma_0}{\sigma} - \alpha(X) \right) \vert X = X_0 \right\} = \mathbb{E}\left\{ \rho \left(\epsilon \frac{\sigma_0}{\sigma} - \alpha(X_0) \right)  \right\} > \mathbb{E} \left\{ \rho\left( \epsilon \frac{\sigma_0}{\sigma} \right) \right\}.
\end{align*}
Hence, since, by (A6), $\Pr(\mathcal{A}_0) >0$, we find
\begin{align*}
M(f, \sigma)  & > \mathbb{E}\left\{ \rho\left( \epsilon \frac{\sigma_0}{\sigma}   \right) \right\} \Pr(\mathcal{A}_0) + \mathbb{E} \left\{ \mathbb{E}\left\{ \rho\left( \epsilon \frac{\sigma_0}{\sigma} \right) \right\} \mathcal{I}_{\mathcal{A}_0^c}(X) \right\}
\\   & = \mathbb{E}\left\{ \rho\left( \epsilon \frac{\sigma_0}{\sigma}   \right) \right\} \Pr(\mathcal{A}_0) + \mathbb{E}\left\{ \rho\left( \epsilon \frac{\sigma_0}{\sigma} \right) \right\} \Pr(\mathcal{A}_0^c) 
\\ & = M(f_0, \sigma),
\end{align*}
where the strict inequality follows from the fact that $\mathcal{A}_0$ has strictly positive probability. 
\end{proof}

\begin{lemma}
\label{Lem2}
Let $\rho$ satisfy (A1) and $K$ satisfy (A7). Then the following uniform law holds
\begin{align*}
\sup_{\sigma>0, f \in \Theta_K}|M_n(f, \sigma) - M(f, \sigma)| \xrightarrow{a.s.} 0,
\end{align*}
as $n \to \infty$.
\end{lemma}
\begin{proof}
The proof may be deduced from the proof of Lemma A.1.2 in \citet{Boente:2020}, we omit the details.
\end{proof}

\begin{lemma}
\label{Lem3}
Suppose that assumptions (A1), (A2), (A4) and (A7) hold. Then, $M(\widehat{f}_n, \sigma_0) \xrightarrow{P} M(f_0, \sigma_0)$, as $n \to \infty$.
\end{lemma}
\begin{proof}
By Lemma~\ref{Lem1}, $f_0$ is the unique minimizer of $M(f, \sigma_0)$ over all $f \in \mathcal{B}([0,1])$ and, by (A7), $\widehat{f}_n \in \mathcal{B}([0,1])$. Therefore,
\begin{align}
\label{eq:A3}
0 \leq M(\widehat{f}_n, \sigma_0)-M(f_0, \sigma_0) = I + II + III,
\end{align}
with
\begin{align*}
I & = M(\widehat{f}_n, \sigma_0) - M_n(\widehat{f}_n, \sigma_0) \\ 
II & = M_n(\widehat{f}_n, \sigma_0) - M_n(\widehat{f}_n, \widehat{\sigma}_n) \\
III & = M_n(\widehat{f}_n, \widehat{\sigma}_n) - M(f_0, \sigma_0)
\end{align*}
Since $\widehat{f}_n \in \Theta_K$ by definition of the estimator,  Lemma~\ref{Lem2} yields
\begin{align}
\label{eq:A4}
|I| \leq  \sup_{\sigma>0, f \in \Theta_K} |M(f, \sigma) - M_n(f, \sigma)| \xrightarrow{a.s.} 0.
\end{align} 
Additionally, a first order Taylor expansion immediately yields
\begin{align}
\label{eq:A5}
|II| \leq \sup_{x \in \mathbb{R}} |x \psi(x)| \frac{|\widehat{\sigma}_n- \sigma_0|}{\widehat{\xi}_n},
\end{align}
where $\widehat{\xi}_n$ is an intermediate value in the linear segment joining $\widehat{\sigma}_n$ and $\sigma_0$. By (A2), $\widehat{\xi}_n >0$ for all large $n$ with high probability and $|\sigma_n- \sigma_0| \xrightarrow{P} 0$ as $n \to \infty$ leading to $|II| \xrightarrow{P} 0$.

To complete the proof we now	 treat $III$. Note that $\widehat{f}_n$ minimizes $M_n(f, \widehat{\sigma}_n) + \lambda \mathcal{J}(f)$ over all $f \in \Theta_K$ and, by construction, $\widetilde{f}_K \in \Theta_K$. Therefore,
\begin{align*}
M_n(\widehat{f}_n, \widehat{\sigma}_n) & \leq M_n(\widehat{f}_n, \widehat{\sigma}_n) + \lambda \mathcal{J}(\widehat{f}_n) \\ 
& \leq M_n(\widetilde{f}_K, \widehat{\sigma}_n) + \lambda \mathcal{J}(\widetilde{f}_K) \\
& = M_n(\widetilde{f}_K, \widehat{\sigma}_n) + o_{\Pr}(1),
\end{align*}
by (A7). Hence,
\begin{align*}
III & \leq M_n(\widetilde{f}_K, \widehat{\sigma}_n) - M(f_0, \sigma_0) + o_{\Pr}(1)
\\ & = \{M_n(\widetilde{f}_K, \widehat{\sigma}_n) - M_n(f_0, \sigma_0)\} + \{M_n(f_0, \sigma_0)- M(f_0, \sigma_0)\} + o_{\Pr}(1).
\end{align*}
By the law of large numbers, $M_n(f_0, \sigma_0) \xrightarrow{P} M(f_0, \sigma_0)$. At the same time, by assumption (A4) and the Schwarz inequality, 
\begin{align*}
|M_n(\widetilde{f}_K, \widehat{\sigma}_n) - M_n(f_0, \sigma_0)| & \leq  |M_n(\widetilde{f}_K, \widehat{\sigma}_n) - M_n(\widetilde{f}_K, \sigma_0)| + |M_n(\widetilde{f}_K, \sigma_0) - M_n(f_0, \sigma_0)|
\\ & \leq \sup_{x \in \mathbb{R}} |x \psi(x)| \frac{|\widehat{\sigma}_n - \sigma_0|}{\widehat{\xi}_n} + \frac{C||\psi||_{\infty}}{\sigma_0}||\widetilde{f}_K-f_0||,
\end{align*}
with probability one, where $\widehat{\xi}_n$ is an intermediate point. By assumptions (A2) and (A7) both of these terms tend to zero in probability and we have thus shown that $III \leq o_{\Pr}(1)$, which in combination with \eqref{eq:A4} and \eqref{eq:A5} now completes the proof.

\end{proof}

\begin{proof}[Proof of Theorem~\ref{Thm:1}]

The first part of the Theorem follows from a simple adaptation of Lemma A.1.4 in \citet{Boente:2020}, because by assumption (A5), $\{f \in \mathcal{B}([0,1]): ||f||_{\mathcal{B}} \leq 1 \}$ is compact in the $|| \cdot ||_{\infty}$-topology. 

The first result of the theorem implies that for every $\epsilon>0$ there exists a $L= L_{\epsilon}$ such that $\Pr(||\widehat{f}_n - f_0||_{\mathcal{B}} > L) < \epsilon$ for all large $n$. Thus, it suffices to restrict attention to the set $\{ ||\widehat{f}_n - f_0||_{\mathcal{B}} \leq L \}$. To prove uniform convergence it suffices to show that
\begin{align}
\label{eq:A6}
\inf_{f \in \mathcal{B}([0,1]): ||f-f_0||_{\mathcal{B}} \leq L, ||f-f_0||_{\infty} \geq \epsilon} M(f, \sigma_0) > M(f_0, \sigma_0).
\end{align}
This is sufficient, because by Lemma~\ref{Lem3}, $M(\widehat{f}_n, \sigma_0) \xrightarrow{P} M(f_0, \sigma_0)$ which implies that with high probability $||\widehat{f}_n-f_0||_{\infty} < \epsilon$ for all large $n$. To establish \eqref{eq:A6}, let $\{f_k\}_k$ denote a minimizing sequence, i.e., a sequence satisfying $||f_k-f_0||_{\mathcal{B}} \leq L$, $||f_k-f_0||_{\infty} \geq \epsilon$ and
\begin{align*}
\lim_{k \to \infty} M(f_k, \sigma_0)  = \inf_{f \in \mathcal{B}([0,1]): ||f-f_0||_{\mathcal{B}} \leq L, ||f-f_0||_{\infty} \geq \epsilon} M(f, \sigma_0).
\end{align*}
Such a sequence exists, because $\rho$ is nonnegative and therefore the infimum is bounded from below by $0$.  By compactness in assumption (A5), there exists a subsequence $f_{k_j}-f_0$ which converges uniformly to a function $f^{\star} \in \mathcal{C}([0,1])$. By continuity of the norm, this implies that $\lim_{j \to \infty} ||f_{k_j}-f_0||_{\infty} = ||f^{\star}||_{\infty}$ and since $||f_k-f_0||_{\infty} \geq \epsilon$ for all $k \in \mathbb{N}$, we must also have $||f^{\star}||_{\infty} \geq \epsilon$. By the bounded convergence theorem it now follows that
\begin{align*}
\inf_{f \in \mathcal{B}([0,1]): ||f-f_0||_{\mathcal{B}} \leq L, ||f-f_0||_{\infty} \geq \epsilon} M(f, \sigma_0) = \lim_{j \to \infty} M(f_{k_j}, \sigma_0) = M(f_0+f^{\star}, \sigma_0)
\end{align*}
and, since, by \eqref{eq:5},
\begin{align*}
||f_0+f^{\star}-f_0||_{\mathcal{B}} = ||f^{\star}||_{\mathcal{B}} \geq c_0^{-1} ||f^{\star}||_{\infty} \geq c_0^{-1} \epsilon > 0,
\end{align*}
where $c_0$ is the embedding constant. It now follows from Lemma~\ref{Lem1} that $M(f_0+f^{\star}, \sigma_0) > M(f_0, \sigma_0)$, concluding the proof.

\end{proof}

We now introduce some useful notation. Let $\mathcal{G}$ denote a class of real-valued functions on $\mathcal{L}^2([0,1]) \times \mathbb{R}$. For $g \in \mathcal{G}$ we define
\begin{align*}
||g||_{\infty} = \sup_{x \in \mathcal{L}^2([0,1]), y \in \mathbb{R}}|g(x,y)|.
\end{align*}
The covering number in this uniform metric, $\mathcal{N}_{\infty}(\epsilon, \mathcal{G})$, is defined as the smallest value of $N \in \mathbb{N}$ such that there exists a sequence $\{g_j\}_{j=1}^N$ with the property that
\begin{align*}
\sup_{g \in \mathcal{G}} \min_{j=1, \ldots, N} ||g-g_j||_{\infty} \leq \epsilon. 
\end{align*}
The corresponding entropy, $\mathcal{H}_{\infty}(\epsilon, \mathcal{G})$, is defined as the logarithm of the covering number, i.e.,  $\mathcal{H}_{\infty}(\epsilon, \mathcal{G})= \log \mathcal{N}_{\infty}(\epsilon, \mathcal{G}) $. 

For a probability measure $\mathbb{P}$ we also define the bracketing number in the $\mathcal{L}^2(\mathbb{P})$-metric, $\mathcal{N}_B(\epsilon, \mathcal{G}, \mathbb{P})$, as the smallest value of $N \in \mathbb{N}$ for which there exist $N$ pairs of functions $\{[g_j^L, g_j^U] \}$ such that $||g_j^U-g_j^L||_{\mathcal{L}^2(\mathbb{P})}\leq \epsilon $ for all $j=1, \ldots, N,$ and such that for each $g \in \mathcal{G}$, there is a $j=j(g) \in \{1, \ldots, N\}$ such that
\begin{align*}
g_j^L(x,y) \leq g(x,y) \leq g_j^U(x,y).
\end{align*}
The corresponding bracketing entropy, $\mathcal{H}_B(\epsilon, \mathcal{G}, \mathbb{P})$, is defined as the logarithm of the bracketing number, i.e.,  $\mathcal{H}_B(\epsilon, \mathcal{G}, \mathbb{P}) = \log  \mathcal{N}_B(\epsilon, \mathcal{G}, \mathbb{P})$.

\begin{lemma}[Bracketing entropy]
\label{Lem4}
 Suppose that (A1) and (A4) hold.  For $(x, y) \in \mathcal{L}^2([0,1]) \times \mathbb{R}$ define
\begin{align*}
g_{f, \sigma}(x,y) = \rho\left( \frac{y-\langle x, f \rangle}{\sigma} \right) - \rho\left( \frac{y-\langle x, \widetilde{f}_K \rangle}{\sigma} \right)
\end{align*}
and the class of functions 
\begin{align*}
\mathcal{G}_{n, c, \delta} = \left\{g_{f, \sigma}(x,y),\ f \in \Theta_K, \ ||f-\widetilde{f}_K|| \leq c,\ |\sigma-\sigma_0| \leq \delta \right\}.
\end{align*}
Let $\mathbb{P}$ denote the probability measured induced by $(X,y)$. Then, there exists a constant $A>0$ depending only on $c$ and $\delta$ such that
\begin{align*}
H_{B} \left(\epsilon, \mathcal{G}_{n, c,\delta}, \mathbb{P} \right) \leq  3K \log\left(1+\frac{A}{\epsilon}\right).
\end{align*}
\end{lemma}
\begin{proof}
Let us begin by observing that, by Lemma 2.1 of \citet{van de Geer:2000}, we have
\begin{align*}
\mathcal{H}_{B} (\epsilon, \mathcal{G}_{n, c,\delta},\mathbb{P} )  \leq \mathcal{H} _{\infty} (\epsilon/2, \mathcal{G}_{n, c,\delta} ),
\end{align*}
so that it suffices to bound the covering number of $\mathcal{G}_{n,c,\delta}$ in the uniform metric. Applying the triangle inequality twice now yields
\begin{align*}
|g_{f_1, \sigma_1}(x,y) - g_{f_2, \sigma_2}(x,y)| & \leq |g_{f_1, \sigma_1}(x,y) - g_{f_1, \sigma_2}(x,y)| + |g_{f_1, \sigma_2}(x,y) - g_{f_2, \sigma_2}(x,y)|
\\ & \leq \frac{2}{\sigma_0-\delta} \sup_{x \in \mathbb{R}} |x \psi(x)|| \sigma_1-\sigma_2| + \frac{C}{\sigma_0-\delta}||f_1-f_2||
\\ & \leq \frac{2}{\sigma_0-\delta} \sup_{x \in \mathbb{R}} |x \psi(x)|| \sigma_1-\sigma_2| + \frac{C}{\sigma_0-\delta}||f_1-\widetilde{f}_K|| + \frac{C}{\sigma_0-\delta}||f_2-\widetilde{f}_K||,
\end{align*}
where we have used (A4). This implies that modulo some constants the covering number in the uniform metric may be bounded by the covering number of a Euclidean ball with radius $\delta$ and the square of the covering number of a set of functions in $\mathcal{L}^2([0,1])$ with radius $c$, viz,
\begin{align*}
\mathcal{N}_{\infty} (\epsilon, \mathcal{G}_{n, c,\delta} ) \leq \mathcal{N}(c_1\epsilon, \mathcal{V}_{\sigma_0}) \times \mathcal{N}^2(c_2 \epsilon, \{f \in \Theta_K: ||f-\widetilde{f}_K|| \leq c \}),
\end{align*}
for $V_{\sigma_0} = [\sigma_0-\delta, \sigma_0+\delta]$, $c_1 = (\sigma_0-\delta)/(8\sup_{x} |x\psi(x)|)$ and $c_2 = (\sigma_0-\delta)/(4C)$. By Lemma 2.5 and Corollary 2.6 of \citet{van de Geer:2000} respectively, these covering numbers may be bounded by
\begin{align*}
\mathcal{N}_{\infty} \left(\epsilon, \mathcal{G}_{n, c,\delta} \right) \leq  \left( \frac{2\sigma_0}{c_1 \epsilon} + 1 \right) \times \left( \frac{4c}{c_2 \epsilon}+1 \right)^{2K}
\leq \left(\frac{A^{\prime}}{\epsilon} +1 \right)^{2K+1},
\end{align*}
for $A^{\prime}= \max\{ 2 \sigma_0/c_1, 4c/c_2 \}$. Now take logarithms, put $A=2 A^{\prime}$ and use that $K \geq 1$. 
\end{proof}

\begin{proof}[Proof of Theorem~\ref{Thm:2}]

To establish Theorem~\ref{Thm:2} we fill in the details of the development in Section \ref{sec:convrate}.
In particular, we first establish \eqref{eq:8}, then we prove \eqref{eq:9} and finally we deduce Theorem~\ref{Thm:2} from \eqref{eq:10}.

First, note that by Theorem~\ref{Thm:1}, $||\widehat{f}_n - f_0||_{\infty} \xrightarrow{\Pr} 0$ and by assumption (A2), $\widehat{\sigma}_n \xrightarrow{\Pr} \sigma_0$. Therefore, we may restrict attention to the set 
\begin{align}
\label{eq:A7}
F_n = \{||\widehat{f}_n - f_0||_{\infty} < \delta \wedge |\widehat{\sigma}_n-\sigma_0| < \delta  \}
\end{align}
for some small $\delta>0$ to be chosen later.

To prove \eqref{eq:8}, it suffices to show that, for all $f \in \Theta_K$ and $\sigma >0$ satisfying $||f-f_0||_{\infty} <\delta$ and $|\sigma-\sigma_0| < \delta$ respectively, we have
\begin{align}
M(f, \sigma) - M(\widetilde{f}_K, \sigma) \geq \eta |\pi(f,\widetilde{f}_K)|^2 - L||\widetilde{f}_K-f_0|| \pi(f, \widetilde{f}_K), 
\label{eq:A7b}
\end{align}
for some $\eta>0$ and $L>0$ with high probability.  To see that this is sufficient, let us assume without loss of generality that $\eta |\pi(\widehat{f}_n,\widetilde{f}_K)|^2 - L||\widetilde{f}_K-f_0|| \pi(\widehat{f}_n, \widetilde{f}_K) > 0$ for all large $n$ (if that were not true for some $n$, then $\eta |\pi(\widehat{f}_n,\widetilde{f}_K)| \leq  L||\widetilde{f}_K-f_0||$ and there is nothing to prove). Then,
\begin{align*}
M(\widehat{f}_n,\widehat{\sigma}_n) - M(\widetilde{f}_K, \widehat{\sigma}_n) & = \frac{M(\widehat{f}_n,\widehat{\sigma}_n) - M(\widetilde{f}_K, \widehat{\sigma}_n)}{\eta |\pi(\widehat{f}_n,\widetilde{f}_K)|^2 - L||\widetilde{f}_K-f_0|| \pi(\widehat{f}_n, \widetilde{f}_K)} \\ & \quad  \times
\{\eta |\pi(\widehat{f}_n,\widetilde{f}_K)|^2 - L||\widetilde{f}_K-f_0|| \pi(\widehat{f}_n, \widetilde{f}_K)\}
\\ & \geq \inf_{\substack{||f-f_0||_{\infty} < \delta, |\sigma- \sigma_0|<\delta \\ \eta |\pi(f,\widetilde{f}_K)|^2 - L||\widetilde{f}_K-f_0|| \pi(f, \widetilde{f}_K)>0}} \frac{M(f,\sigma)-M(\widetilde{f}_K, \sigma)}{\eta |\pi(f,\widetilde{f}_K)|^2 - L||\widetilde{f}_K-f_0|| \pi(f, \widetilde{f}_K)}
\\ & \quad \times
\{\eta |\pi(\widehat{f}_n,\widetilde{f}_K)|^2 - L||\widetilde{f}_K-f_0|| \pi(\widehat{f}_n, \widetilde{f}_K)\}
\\ & \geq \eta |\pi(\widehat{f}_n,\widetilde{f}_K)|^2 - L||\widetilde{f}_K-f_0|| \pi(\widehat{f}_n, \widetilde{f}_K),
\end{align*}
since the infimum is $\geq 1$ according to (\ref{eq:A7b}).  

We thus have to prove  inequality (\ref{eq:A7b}) to establish (\ref{eq:8}). First, write
\begin{align*}
Y - \langle X, f \rangle = \sigma_0 \epsilon + R + \langle X, \widetilde{f}_{K} - f \rangle,
\end{align*}
where $R = \langle X, f_0 - \widetilde{f}_K \rangle $. A first order Taylor expansion with Lagrange remainder yields
\begin{align}
\label{eq:A8}
M(f, \sigma) - M(\widetilde{f}_K, \sigma) 
& = \mathbb{E} \left\{ \psi\left(\frac{\sigma_0 \epsilon + R}{\sigma} \right) \frac{\langle X, \widetilde{f}_K - f \rangle }{\sigma} \right\}  \nonumber \\
&  \quad +\frac{1}{2} \mathbb{E}\left\{ \psi^{\prime}\left(\frac{\sigma_0 \epsilon + R + \xi}{\sigma} \right) \left| \frac{\langle X, \widetilde{f}_K-f \rangle}{\sigma} \right|^2  \right\},
\end{align}
for some random variable $\xi$ satisfying $|\xi| \leq |\langle X, \widetilde{f}_K-f \rangle|$. Applying the mean-value theorem on the first term of the rhs of \eqref{eq:A8} we also find that there exists a random variable $\chi$ such that $|\chi|\leq |R|$ and
\begin{align}
\mathbb{E} \left\{ \psi\left(\frac{\sigma_0 \epsilon + R}{\sigma} \right) \frac{\langle X, \widetilde{f}_K -f \rangle }{\sigma} \right\} &= \mathbb{E} \left\{ \psi\left(\frac{\sigma_0 \epsilon}{\sigma} \right) \frac{\langle X, \widetilde{f}_K -f \rangle }{\sigma} \right\} \nonumber
\\ & \quad + \mathbb{E}\left\{ R \psi^{\prime}\left(\frac{\sigma_0 \epsilon + \chi}{\sigma} \right) \frac{\langle X, \widetilde{f}_K -f \rangle }{\sigma} \right\} \nonumber
\\ & = \mathbb{E}\left\{ R \psi^{\prime}\left(\frac{\sigma_0 \epsilon + \chi}{\sigma} \right) \frac{\langle X, \widetilde{f}_K -f \rangle }{\sigma} \right\}.
\label{eq:A9}
\end{align}
To see why the first term vanishes, note that $\mathbb{E}\{\psi(\sigma_0 \epsilon/ \sigma \} = 0$ for any $\sigma>0$ because Lemma~\ref{Lem1} shows that $f_0$ minimizes $M(f, \sigma)$. For the remaining term in \eqref{eq:A9}, by noting again that $\sigma > \sigma_0-\delta$ we obtain
\begin{align*}
\left|\mathbb{E}\left\{ R \psi^{\prime}\left(\frac{\sigma_0 \epsilon + \chi}{\sigma} \right) \frac{\langle X, \widetilde{f}_K - f \rangle }{\sigma} \right\} \right| &\leq (\sigma_0-\delta)^{-1}||\psi^{\prime}||_{\infty} \mathbb{E}\{|\langle X, \widetilde{f}_K-f_0 \rangle \langle X, \widetilde{f}_K-f \rangle| \}
\\ & \leq C (\sigma_0-\delta)^{-1}||\psi^{\prime}||_{\infty} ||\widetilde{f}_K-f_0|| \left[\mathbb{E}\{ | \langle X, f-\widetilde{f}_K \rangle|^2 \} \right]^{1/2}
\\ & = L_{\delta} ||\widetilde{f}_K-f_0|| \pi(f,\widetilde{f}_K),
\end{align*}
for $L_{\delta} =  C (\sigma_0-\delta)^{-1}||\psi^{\prime}||_{\infty}$. This is exactly the second term in the rhs of \eqref{eq:8}.

The last part of the proof establishes a strictly positive lower bound on the second term of \eqref{eq:A8} involving $|\pi(f, \widetilde{f}_K)|^2$. Note that for all $f \in \Theta_K$ satisfying $||f-f_0||_{\infty} < \delta$ we have
\begin{align*}
||\widetilde{f}_K - f||_{\infty} \leq ||\widetilde{f}_K - f_0||_{\infty} + ||f - f_0||_{\infty} < 2 \delta,
\end{align*}
for all large $n$, by virtue of (A7). Since $X$ is bounded by (A4), for all large $n$, $|\xi| \leq 2C \delta$.  Assumption (A7) in combination with (A4) also implies $|R| \leq C\delta $ for every $\delta$, for sufficiently large $n$. By (A1) $\psi^{\prime}$ is continuous and bounded, and by (A4), $\mathbb{E}\{\psi^{\prime}(\epsilon) \}>0$. Hence, $m(t, \sigma) := \mathbb{E}\{\psi((\sigma_0 \epsilon+t)/\sigma \}$ is continuous at $(0, \sigma_0)$ and $m(0, \sigma_0) = \mathbb{E}\{\psi^{\prime}(\epsilon) \}>0 $. This observation now leads to
\begin{align*}
\inf_{|b| < 3C\delta, |\sigma-\sigma_0| < \delta}\mathbb{E}\left\{ \psi^{\prime}\left(\frac{\sigma_0 \epsilon + b}{\sigma} \right) \right\} \geq \frac{\mathbb{E}\{\psi^{\prime}(\epsilon) \}}{2}>0,
\end{align*}
for all sufficiently small $\delta>0$. Setting $\eta = (\sigma_0+\delta)^{-2} \mathbb{E}\{\psi^{\prime}(\epsilon) \}/4$, we finally have
\begin{align*}
\frac{1}{2} \mathbb{E}\left\{ \psi^{\prime}\left(\frac{\sigma_0 \epsilon + R + \xi}{\sigma} \right) \left| \frac{\langle X, \widetilde{f}_K-f \rangle}{\sigma} \right|^2  \right\} &\geq \eta \mathbb{E}\{ |\langle X, f-\widetilde{f}_K \rangle|^2 \}
\\ & = \eta |\pi(f, \widetilde{f}_K)|^2,
\end{align*}
for all large $n$, completing the first part of the proof.

The second step in our proof is the establishment of \eqref{eq:9}. Recall that by assumption (A2) and Theorem 1, we may restrict attention to the set $F_n$ in \eqref{eq:A7}. As previously remarked, in this set we also have $||\widehat{f}_n - \widetilde{f}_K||_{\infty} \leq 2 \delta$ for all large $n$. It then also follows that $||\widehat{f}_n - \widetilde{f}_K|| \leq 2 \delta$ because the uniform norm dominates the $\mathcal{L}^2([0,1])$-norm. Thus, in the notation of Section~\ref{sec:3},
\begin{align}
\label{eq:A10}
\left|\frac{U_n(\widehat{f}_n, \widetilde{f}_K, \widehat{\sigma}_n)}{\gamma_n \pi(\widehat{f}_n, \widetilde{f}_K) \vee \gamma_n^2} \right| \leq \sup_{\substack{ f \in \Theta_K : ||f-\widetilde{f}_K|| \leq 2 \delta \\ |\sigma-\sigma_0| \leq \delta }} \left| \frac{U_n(f, \widetilde{f}_K, \sigma)}{\gamma_n \pi(f, \widetilde{f}_K) \vee \gamma_n^2} \right|,
\end{align}
for all large $n$.  To prove \eqref{eq:9} it suffices to show that the random variable in the rhs of \eqref{eq:A10} is bounded in probability. For convenience, let $\Phi_{K, \delta} = \{f \in \Theta_K: ||f-\widetilde{f}_K|| \leq 2\delta\}$ and $V_{\sigma_0, \delta} = [\sigma_0-\delta, \sigma_0+\delta]$, then we equivalently need to show that 
\begin{align}
\label{eq:A11}
\lim_{T \to \infty} \limsup_{n \to \infty} \Pr\left(\sup_{\substack{f \in \Phi_{K, \delta}, \sigma \in V_{\sigma_0,\delta} \\ \pi(f, \widetilde{f}_K) \leq \gamma_n }} \left|U_n(f, \widetilde{f}_K, \sigma)\right| \geq T \gamma_n^2 \right) = 0,
\end{align}
as well as
\begin{align}
\label{eq:A12}
\lim_{T \to \infty} \limsup_{n \to \infty} \Pr\left( \sup_{\substack{f \in \Phi_{K,\delta}, \sigma \in V_{\sigma_0,\delta} \\ \pi(f, \widetilde{f}_K) > \gamma_n }} \left|\frac{U_n(f, \widetilde{f}_K, \sigma)}{\pi(f, \widetilde{f}_K)}  \right|  \geq T \gamma_n \right) = 0.
\end{align}

First, observe that for all $\epsilon>0$ sufficiently small, say $\epsilon \leq \epsilon_0$, there exists a constant $B>0$ such that
\begin{align}
\label{eq:A13}
\int_0^{\epsilon} \log^{1/2}\left(1+\frac{1}{u}\right) du \leq B \epsilon \log^{1/2}\left( \frac{1}{\epsilon} \right).
\end{align}
This inequality will be useful in the derivation of both \eqref{eq:A11} and \eqref{eq:A12}. To show \eqref{eq:A11}, we aim to apply Theorem 5.11 of \citet{van de Geer:2000} on this mean-centered process. Let us rewrite $U_n(f, \widetilde{f}_K, \sigma)$ in terms of the empirical process:
\begin{align}
\label{A14}
U_n(f, \widetilde{f}_K, \sigma) = n^{-1/2} v_n(g_{f, \sigma}), 
\end{align}
where, as in Lemma~\ref{Lem4},
\begin{align*}
g_{f, \sigma}(X,y) = \rho\left(\frac{y - \langle X, f \rangle}{\sigma}\right) - \rho\left(\frac{y - \langle X, \widetilde{f}_K \rangle}{\sigma}\right),
\end{align*}
and $v_n(g_{f,\sigma}) = \int g_{f,\sigma} d(\mathbb{P}_n-\mathbb{P})$, with $\mathbb{P}_n$ the empirical measure. By assumption (A1),
\begin{align*}
\sup_{\substack{f \in \Phi_{K,\delta}, \sigma \in V_{\sigma_0,\delta} \\ \pi(f, \widetilde{f}_K) \leq \gamma_n }}|g_{f,\sigma}| \leq 2,
\end{align*} 
and 
\begin{align*}
\{g_{f, \sigma}, f \in \Phi_{K,\delta}, \sigma \in V_{\sigma_0, \delta}, \pi(f, \widetilde{f}_K) \leq \gamma_n  \} \subset \mathcal{G}_{n,2 \delta, \delta}.
\end{align*}
Thus, by Lemma 5.10 of \citet{van de Geer:2000}, the generalized entropy with bracketing in the Bernstein norm $\mathcal{H}_{B,8}(\epsilon, \mathcal{G}_{n,2\delta,\delta}, \mathbb{P})$ may be bounded by the bracketing entropy, i.e.,
\begin{align}
\label{eq:A15}
\mathcal{H}_{B,8}\left(\epsilon, \mathcal{G}_{n,2\delta,\delta}, \mathbb{P}\right) \leq \mathcal{H}_{B}\left(\epsilon/\sqrt{2}, \mathcal{G}_{n,2\delta,\delta}, \mathbb{P}\right) \leq 3 K \log\left(1+\frac{\sqrt{2}A}{\epsilon}\right),
\end{align}
where the last inequality follows from Lemma~\ref{Lem4}. Furthermore, by (A1) we have
\begin{align*}
|g_{f, \sigma}(X,y)| \leq \sigma^{-1}||\psi||_{\infty} |\langle X, \widetilde{f}_K-f \rangle|.
\end{align*}
Consequently, by definition of $\pi(f,g)$,
\begin{align*}
\sup_{\substack{f \in \Phi_{K,\delta}, \sigma \in V_{\sigma_0,\delta} \\ \pi(f, \widetilde{f}_K) \leq \gamma_n }} \mathbb{E} \{ | g_{f, \sigma}(X,y)|^2 \} \leq \frac{||\psi||_{\infty}^2}{(\sigma_0-\delta)^2}  \gamma_n^2
\end{align*}
It follows by Lemma 5.8 of \citet{van de Geer:2000} that we may take $R = c^{\prime} \gamma_n $ with $c^{\prime} = ||\psi||_{\infty}/(\sigma_0-\delta)$ in Theorem 5.11 of \citet{van de Geer:2000}. We proceed to check the conditions of the theorem. By (A7) we have that $\gamma_n \to 0$ as $n \to \infty$. Hence, by a change of variables and \eqref{eq:A15}, we find
\begin{align*}
\int_{0}^R \mathcal{H}_{B,8}^{1/2}(u, \mathcal{G}_{n,2\delta,\delta}, \mathbb{P}) du \leq C_0 K^{1/2} \gamma_n \log^{1/2} n,
\end{align*}
for some $C_0>0$. By taking $T = C_0$ and $C_1 = 8 C_0/c_0^2$ in the theorem, it may be seen that conditions (5.31)--(5.34) in \citet{van de Geer:2000} are satisfied for $\alpha = T n^{1/2}\gamma_n^2$ and sufficiently large $C_0$. Thus, there exists a universal constant $C>0$ such that
\begin{align*}
\Pr\left(\sup_{\substack{f \in \Phi_K, \sigma \in V_{\sigma} \\ \pi(f, \widetilde{f}_K) \leq \gamma_n }} \left|U_n(f, \widetilde{f}_K, \sigma)\right| \geq T \gamma_n^2 \right) & = \Pr\left( \sup_{\substack{f \in \Phi_K, \sigma \in V_{\sigma} \\ \pi(f, \widetilde{f}_K) \leq \gamma_n }} \left|v_n(g_{f,s})\right| \geq T n^{1/2}\gamma_n^2\right) \\ & \leq C \exp\left[ - \frac{T^2 K \log n}{C^2(C_1+1)} \right].
\end{align*}
Since $K \to \infty$ as $n \to \infty$, the exponential tends to zero as $n \to \infty$ and since this holds for all $T$ sufficiently large, \eqref{eq:A11} now follows from Theorem 5.11 of \citet{van de Geer:2000}.

To show \eqref{eq:A12} we modify the peeling argument given in Lemma 5.13 of \citep{van de Geer:2000}. First, note that, by (A4), $\pi(f, \widetilde{f}_K) \leq C ||f-\widetilde{f}_K||$ and $||f - \widetilde{f}_K|| \leq 2 \delta$ for all $f \in \Phi_{K,\delta}$. By choosing $\delta \leq \epsilon_0/(2Cc^{\prime})$, with $\epsilon_0$ determined in \eqref{eq:A13} and $c^{\prime} = ||\psi||_{\infty}/(\sigma_0-\delta)$, we may assume without loss of generality that $\pi(f, \widetilde{f}_K) \leq \epsilon_0/c^{\prime}$ for all $f \in \Phi_{K, \delta}$. Thus, to prove \eqref{eq:A12}, it suffices to prove
\begin{align}
\label{eq:A16}
\lim_{T \to \infty} \limsup_{n \to \infty} \Pr\left( \sup_{\substack{f \in \Phi_{K,\delta}, \sigma \in V_{\sigma_0} \\ \gamma_n< \pi(f, \widetilde{f}_K) \leq \epsilon_0/c^{\prime} }} \left|\frac{U_n(f, \widetilde{f}_K, \sigma)}{\pi(f, \widetilde{f}_K)}  \right|  \geq T \gamma_n \right) = 0.
\end{align}
Now, let $S = \min\{s >1 : 2^{-s} \epsilon_0/c^{\prime} < \gamma_n \}$. Since, by assumption (A7), $K \asymp n^{\beta}$ for $\beta \in (0,1)$ we clearly have $S \leq [c\log_2 n+1]$ for some $c>0$. Using Boole's inequality we obtain
\begin{align*}
\Pr\left( \sup_{\substack{f \in \Phi_{K,\delta}, \sigma \in V_{\sigma_0,\delta} \\ \gamma_n< \pi(f, \widetilde{f}_K) \leq \epsilon_0/c^{\prime} }} \left|\frac{U_n(f, \widetilde{f}_K, \sigma)}{\pi(f, \widetilde{f}_K)}  \right|  \geq T \gamma_n \right) & \leq \sum_{s=1}^{S} \Pr \left( \sup_{\substack{f \in \Phi_{K,\delta}, \sigma \in V_{\sigma_0,\delta} \\ 2^{-s} \epsilon_0/c^{\prime} < \pi(f, \widetilde{f}_K) \leq 2^{-s+1} \epsilon_0/c^{\prime} }} \left|\frac{U_n(f, \widetilde{f}_K, \sigma)}{\pi(f, \widetilde{f}_K)}  \right|  \geq T \gamma_n \right)
\\ &\leq \sum_{s=1}^{S} \Pr \left( \sup_{\substack{f \in \Phi_{K,\delta}, \sigma \in V_{\sigma_0,\delta} \\ \pi(f, \widetilde{f}_K) \leq 2^{-s+1} \epsilon_0/c^{\prime}	 }} \left|U_n(f, \widetilde{f}_K, \sigma)  \right|  \geq T 2^{-s} \frac{\epsilon_0}{c^{\prime}}\gamma_n \right).
\end{align*}
We bound each one of these summands through individual application of Theorem 5.11 of \citet{van de Geer:2000} (see also the proof of \eqref{eq:A11}). Rewriting in terms of the empirical process we have
\begin{align*}
\Pr \left( \sup_{\substack{f \in \Phi_{K,\delta}, \sigma \in V_{\sigma_0,\delta} \\ \pi(f, \widetilde{f}_K) \leq 2^{-s+1} \epsilon_0 /c^{\prime} }} \left|U_n(f, \widetilde{f}_K, \sigma)  \right|  \geq T 2^{-s} \frac{\epsilon_0}{c^{\prime}}\gamma_n \right) = \Pr \left( \sup_{\substack{f \in \Phi_{K,\delta}, \sigma \in V_{\sigma_0,\delta} \\ \pi(f, \widetilde{f}_K) \leq 2^{-s+1} \epsilon_0/c^{\prime} }} |v_n(g_{f,\sigma})|  \geq T 2^{-s} \frac{\epsilon_0}{c^{\prime}} n^{1/2}\gamma_n \right)
\end{align*}
Clearly, $2^{-s+1} c^{\prime} \epsilon_0/c^{\prime} \leq \epsilon_0$ for all $1 \leq s \leq S$. Hence, for all sufficiently large $C_0$ the bracketing integral for each one of these classes may be bounded by
\begin{align*}
\int_{0}^{2^{-s+1}\epsilon_0}\mathcal{H}_{B,8}^{1/2}(u, \mathcal{G}_{n,2\delta,\delta}, \mathbb{P}) du &\leq C_0 K^{1/2} 2^{-s+1} \epsilon_0 \log n,
\end{align*}
for all large $n$, by the construction of $S$, i.e., $2^{S} \leq 2c n $ for large $n$. The conditions of Theorem 5.11 in \citet{van de Geer:2000} are satisfied for sufficiently large $C_0$,  $C_1 = 8C_0 c^{\prime}$ and $T = C_0$ since by definition of $S$, we have $\gamma_n \leq 2^{-s+1} \epsilon_0/c^{\prime}$ for all $1 \leq s \leq S$. Thus, this theorem yields
\begin{align*}
\Pr \left( \sup_{\substack{f \in \Phi_{K,\delta}, \sigma \in V_{\sigma_0,\delta} \\ \pi(f, \widetilde{f}_K) \leq 2^{-s+1} \epsilon_0 /c^{\prime} }} \left|U_n(f, \widetilde{f}_K, \sigma)  \right|  \geq T 2^{-s} \frac{\epsilon_0}{c^{\prime}}\gamma_n \right) \leq C \exp\left[ - \frac{T^2 K \log n}{4 C^2(C_1+1)  |c^{\prime}|^2 } \right],
\end{align*}
for the same universal constant $C>0$. None of these terms depend on $s$, hence after summing over $s \in \{1, \ldots, S\}$ and recalling that $S \leq [c\log_2 n+1]$ we obtain
\begin{align*}
\Pr\left( \sup_{\substack{f \in \Phi_{K,\delta}, \sigma \in V_{\sigma_0,\delta} \\ \gamma_n< \pi(f, \widetilde{f}_K) \leq \epsilon_0/c^{\prime} }} \left|\frac{U_n(f, \widetilde{f}_K, \sigma)}{\pi(f, \widetilde{f}_K)}  \right|  \geq T \gamma_n \right) & \leq C^{\prime} [c\log_2 n+1] \exp\left[ - \frac{T^2 K \log n}{4 C^2(C_1+1)  |c^{\prime}|^2 } \right],
\end{align*}
for some $C^{\prime}>0$ and all large $n$. We have thus established \eqref{eq:A12} and part (ii) now follows. 

To complete the proof we now deduce Theorem~\ref{Thm:2} from \eqref{eq:10}. First, note that by (A4) we have
\begin{align}
\label{eq:A17}
|\pi(\widehat{f}_n, f_0)|^2 \leq 2 |\pi(\widehat{f}_n, \widetilde{f}_K)|^2 + 2 |\pi(\widetilde{f}_K), f_0)|^2 \leq 2 |\pi(\widehat{f}_n, \widetilde{f}_K)|^2 + 2 C^2||\widetilde{f}_K-f_0||^2.
\end{align}
Hence, we need to study $|\pi(\widehat{f}_n, \widetilde{f}_K)|^2$. We only have to handle the case  $\gamma_n \pi(\widehat{f}_n, \widetilde{f}_K) > \gamma_n^2$, or, equivalently $\pi(\widehat{f}_n, \widetilde{f}_K) > \gamma_n$, since for $\gamma_n \pi(\widehat{f}_n, \widetilde{f}_K) \leq \gamma_n^2$, the theorem clearly holds. From parts (i) and (ii) we have
\begin{align*}
\eta |\pi(\widehat{f}_n, \widetilde{f}_K)|^2  & \leq U_n(\widetilde{f}_K,\widehat{f}_n, \widehat{\sigma}_n) + L ||\widetilde{f}_K-f_0|| \pi(\widehat{f}_n, \widetilde{f}_K)  + \lambda \mathcal{J}(\widetilde{f}_K)
\\ & = O_{\Pr}(1)\gamma_n \pi(\widehat{f}_n, \widetilde{f}_K)  + L||\widetilde{f}_K-f_0|| \pi(\widehat{f}_n, \widetilde{f}_K) + \lambda \mathcal{J}(\widetilde{f}_K).
\end{align*}
Equivalently, since $\eta>0$,
\begin{align*}
|\pi(\widehat{f}_n, \widetilde{f}_K)|^2 \leq O_{\Pr}(1)\gamma_n \pi(\widehat{f}_n, \widetilde{f}_K)  + O_{\Pr}(1)||\widetilde{f}_K-f_0|| \pi(\widehat{f}_n, \widetilde{f}_K) + \lambda \mathcal{J}(\widetilde{f}_K)/\eta.
\end{align*}
Now, this is an inequality of the form $x_0^2 \leq bx_0 + c$ with $x_0 = \pi(\widehat{f}_n, \widetilde{f}_K) \geq  0, b = O_{\Pr}(1)(||\widetilde{f}_K-f_0||+\gamma_n)$ and $c =  \lambda \mathcal{J}(\widetilde{f}_K)/\eta$. This means that $x_0$ must be less than or equal to the positive root of $x^2-bx-c=0$, that is,
\begin{align*}
0 \leq x_0 \leq \frac{b+\sqrt{b^2+4c}}{2} \leq b + \sqrt{c},
\end{align*}
and after substituting the expressions of $x_0$, $b$ and $c$, we obtain
\begin{align*}
\pi(\widehat{f}_n, \widetilde{f}_K) \leq O_{\Pr}(1) (||\widetilde{f}_K-f_0|| +\gamma_n) + O_{\Pr}\left(\sqrt{\lambda \mathcal{J}(\widetilde{f}_K)}) \right).
\end{align*}
Squaring and using the inequality $(x+y)^2 \leq 2x^2+2y^2$ twice yields 
\begin{align*}
|\pi(\widehat{f}_n, \widetilde{f}_K)|^2 \leq O_{\Pr}(1) \gamma_n^2 + O_{\Pr}(1) ||\widetilde{f}_K-f_0||^2+ O_{\Pr}(1)\lambda \mathcal{J}(\widetilde{f}_K).
\end{align*}
The result of the theorem now follows easily from \eqref{eq:A17} which completes the proof.
\end{proof}

\section*{References}

\begingroup
\renewcommand{\section}[2]{}%

\endgroup
\end{document}